\documentclass[final,5p,times,twocolumn,sort&compress]{elsarticle}
\usepackage{amssymb}
\usepackage{amsmath}
\pdfoutput=1

\usepackage{cmap}
\usepackage[T2A]{fontenc}

\usepackage{hyperref}
\hypersetup{
	colorlinks=true,
}

\makeatletter
\def\ps@pprintTitle{%
	\let\@oddhead\@empty
	\let\@evenhead\@empty
	\let\@oddfoot\@empty
	\let\@evenfoot\@oddfoot
}
\makeatother

\usepackage{framed}
\usepackage{multicol}

\usepackage{nomencl}
\makenomenclature
\renewcommand*\nompreamble{\begin{multicols}{2}}
	\renewcommand*\nompostamble{\end{multicols}}

\usepackage{natbib}

\usepackage{graphicx}
\usepackage{epstopdf}
\usepackage[skip=-2pt]{subcaption}

\usepackage{tikz}

\usepackage[absolute,overlay]{textpos}

\definecolor{shadecolor}{RGB}{220,220,220}

\usepackage{fixltx2e}

\newcommand\M{\mathrm{M}}
\newcommand\Rey{\mathrm{Re}}
\newcommand\Pran{\mathrm{Pr}}

\DeclareMathOperator{\sign}{sgn}

\renewcommand{\vec}{\boldsymbol}
\newcommand{\mat}{\boldsymbol}

\begin{document}
	\begin{textblock*}{18.4cm}(1.3cm,1.0cm)
		\noindent\footnotesize\colorbox{shadecolor}
		{\parbox{\dimexpr\textwidth-2\fboxsep\relax}{A.I. Aleksyuk, Influence of vortex street structure on the efficiency of energy separation, Int. J. Heat Mass Transfer 135 (2019) 284-293.}}
	\end{textblock*}
	\sloppy
	\begin{frontmatter}
		\title{Influence of Vortex Street Structure on the Efficiency of Energy Separation}		
		\author[MSU,IWP]{Andrey~I.~Aleksyuk}
		\ead{aleksyuk@mech.math.msu.su}
		\address[MSU]{Faculty of Mechanics and Mathematics, Lomonosov Moscow State University, Moscow 119991, Russian Federation}
		\address[IWP]{Water Problems Institute, Russian Academy of Sciences, Moscow 119333, Russian Federation}
		
		\begin{abstract}
			Regions with reduced and increased values of  total enthalpy are observed in a time-averaged flow behind a bluff body. This energy redistribution takes place both in the vortex formation region and in the developed vortex wake. The present paper focuses on studying the effect of the structure of a vortex street on the intensity of energy redistribution. Two approaches are used. The first one is direct numerical simulation of the flow behind a transversely oscillating cylinder, which is known for a variety of vortex patterns in the wake. The simulations are based on a finite element solution of the Navier-Stokes equations for a compressible perfect viscous gas. The second approach is based on simplified point vortex models for infinite periodic vortex streets, which contain a finite number of vortex chains in equilibrium. It turns out that these simple models make it possible to obtain satisfactory qualitative results, particularly if a more precise approximation of velocity fields in the vortex cores (Rankine vortices) is implemented. It is shown that the effect of energy redistribution significantly depends on the vortex structure, namely the mutual arrangement of the vortices and their intensities. The estimates of the energy separation efficiency in the time-averaged flow are obtained for the general case of an arbitrary number of chains. A  more detailed analysis is performed for vortex streets with 2, 3, and 4 vortex chains.
		\end{abstract}
		
	\end{frontmatter}

	\section{Introduction}
	For a viscous-gas flow around a cylinder the phenomenon of energy
	separation is manifested in the appearance of regions with
	increased and reduced values of total enthalpy in the wake. Such
	energy redistribution is observed in both instantaneous and
	time-averaged flow patterns. This problem has been discussed in
	quite a few studies
	\cite{KurosakaEnergyseparationvortex1987,RyanExperimentsaerodynamiccooling1951,ThomannMeasurementsRecoveryTemperature1959,
		EckertCrosstransportenergy1987,NgTimeresolvedmeasurementstotal1990,KulkarniEnergyseparationwake2009,AleksyukDirectnumericalsimulation2018}.
	The interest in the phenomenon of energy separation can be
	partially explained by the attempts to make energy separation
	devices more efficient (see, for example, the Ranque-Hilsch vortex
	tube and the Leontiev tube \cite{Eiamsa-ardReviewRanqueHilsch2008,
		LeontevGasdynamicmethod1997,LeontievExperimentalinvestigationmachinefree2017}).
	Another reason for the increased attention to the problem of
	energy separation in the wake behind bluff bodies is the study of
	the Eckert-Weise effect, which is manifested in low recovery
	temperature at the rear part of a thermally insulated cylinder
	\cite{EckertMessungenTemperaturverteilungauf1942}.
	
	The lowest values of time-averaged total enthalpy are observed in
	the vortex formation region and in the developed wake near its
	centerline \cite{KurosakaEnergyseparationvortex1987,
		AleksyukDirectnumericalsimulation2018,goldstein_energy_2008}. From
	the energy conservation law, it follows  that in a fluid particle
	the total enthalpy can change due to the action of three
	mechanisms: the time-variation of pressure at a given point in space; the work of viscous forces; and the heat release due to the
	thermal conduction effect. The action of all these mechanisms
	should be taken into account in the vortex formation region;
	however, in the developed wake it is possible to consider only the
	first mechanism, related to pressure variation
	\cite{KurosakaEnergyseparationvortex1987,AleksyukDirectnumericalsimulation2018}.

	\begin{table*}[!t]
		\begin{framed}
			\nomenclature[A]{$t$}{time}
			\nomenclature[A]{$(x_1, x_2)$}{vector of Cartesian coordinates fixed to a cylinder}
			\nomenclature[A]{$(x', y')$}{vector of Cartesian coordinates, in which the flow at infinity is at rest}
			\nomenclature[A]{$(x'', y'')$}{vector of Cartesian coordinates fixed to moving vortices}
			\nomenclature[A]{$(x, y)$}{vector of Cartesian coordinates of the inertial reference frame, the free stream velocity is $(1,0)$}
			\nomenclature[A]{$z_k$}{complex coordinates of base potential vortex for chain number $k$}
			\nomenclature[A]{$\vec{n}$}{unit normal vector}
			
			\nomenclature[A]{$l$}{period of the idealized vortex street along the $x$ axis}
			\nomenclature[A]{$N$}{number of vortex chains forming the idealized vortex street}
			\nomenclature[A]{$i$}{unit imaginary number}
			\nomenclature[A]{$E$}{efficiency of energy separation}
			
			\nomenclature[A]{$U$}{$x$-component of vortex velocity in coordinates $(x',y')$}
			\nomenclature[A]{$y_c$}{$y$-coordinate of the cylinder center}
			\nomenclature[A]{$v_c$}{transverse velocity of the cylinder}
			\nomenclature[A]{$p$}{pressure}
			\nomenclature[A]{$(u_1,u_2)$}{velocity vector in coordinates $(x_1,x_2)$}
			\nomenclature[A]{$(u,v)$}{velocity vector in coordinates $(x,y)$}
			\nomenclature[A]{$(u',v')$}{velocity vector in coordinates $(x',y')$}
			\nomenclature[A]{$(u'',v'')$}{velocity vector in coordinates $(x'',y'')$}
			\nomenclature[A]{$T$}{temperature}
			\nomenclature[A]{$i_0$}{total enthalpy}
			\nomenclature[A]{$I_0$}{normalized total enthalpy}
			\nomenclature[A]{$C_L$}{lift coefficient}
			\nomenclature[A]{$c_V$, $c_p$}{specific heats at constant volume and pressure}
			
			\nomenclature[A]{$\mathcal{P},\mathcal{A},\mathcal{Q}$}{contributions of non-stationarity, viscous forces, and thermal conductivity
				in the total-enthalpy variation rate}
			
			\nomenclature[A]{$f'_0$}{vortex shedding frequency for a fixed cylinder}
			\nomenclature[A]{$A$}{amplitude of cylinder oscillations}
			\nomenclature[A]{$F$}{normalized (to $f'_0$) frequency of cylinder oscillations}
			
			\nomenclature[A]{$\Rey$}{Reynolds number}
			\nomenclature[A]{$\M$}{Mach number}
			\nomenclature[A]{$\Pran$}{Prandtl number}
			
			\nomenclature[A]{$d$}{cylinder diameter}
			
			\nomenclature[A]{$E$}{efficiency of energy separation}
			
			\nomenclature[A]{$X_{in}, X_{out}, Y$}{distance from the center of the cylinder to the inlet, outlet, and
				side boundaries of the corresponding subdomains}
			
			\nomenclature[B]{$\Gamma_k$}{circulation of vortices in the chain number $k$}
			\nomenclature[B]{$\Gamma$}{sum of absolute values of circulations $|\Gamma_k|$}
			\nomenclature[B]{$\alpha_k$}{normalized circulations $\Gamma_k/\Gamma$}
			\nomenclature[B]{$\beta$}{ratio $l U/\Gamma$}
			\nomenclature[B]{$\gamma$}{specific-heat ratio}
			\nomenclature[B]{$\varepsilon$}{energy}
			\nomenclature[B]{$\mu$}{viscosity coefficient}
			\nomenclature[B]{$\rho$}{density}
			\nomenclature[B]{$\mat{\tau}$}{viscous stress tensor}
			\nomenclature[B]{$\Delta$}{approximate size of triangular elements of the mesh}
			\nomenclature[B]{$\omega$}{vorticity}
			\nomenclature[B]{$\kappa$}{thermal conductivity}
			
			\nomenclature[C]{$0$}{stagnation (or total) parameters}
			\nomenclature[C]{$\infty$}{free-stream  parameters}
			\nomenclature[C]{$(\cdot)_{,t}$}{time derivatives}
			\nomenclature[C]{$(\cdot)_{,i}$}{coordinate derivatives, $i=1,2$ corresponds to $x,y$}
			
			\nomenclature[D]{$*$}{transposition}
			\nomenclature[D]{$'$}{dimensional parameters}
			\nomenclature[D]{$\overline{(\cdot)}$}{time-averaged value}
			
			\printnomenclature[2 cm]
		\end{framed}
	\end{table*}
	
	Since in the formed vortex wake the action of the mechanism
	related to pressure variation is most significant, the process of
	energy redistribution can be approximately described by the
	equation:
	\begin{equation*}
	\frac{D i_0}{Dt}=\frac{1}{\rho}\frac{\partial p}{\partial t},
	\end{equation*}
	where $i_0$, $p$, $\rho$, and $t$ are total enthalpy, pressure,
	density, and time, respectively. In
	\cite{KurosakaEnergyseparationvortex1987}, the authors explained
	how particles moving inside and outside the wake form a pattern
	with reduced values of $i_0$ near the wake centerline. The main idea
	is that since the pressure inside the vortices is less than that
	in the surrounding fluid, $i_0$ decreases in the fluid particles
	moving in front of the vortex and increases in the particles
	moving behind the vortex. Due to the vortex street structure,
	fluid particles move inside/outside the wake in front of/behind
	the vortices. The kinematical explanation was suggested based on a
	modified classical K\'arm\'an vortex street model (the Rankine
	vortices were used instead of potential vortices). It
	was shown, that the minimum of time-averaged total enthalpy satisfies
	the equation
	\begin{equation*}
	\overline{i_0}-i_{0\infty}=-\frac{\Gamma_0}{l}(1+U),
	\end{equation*}
	where $i_{0\infty}$ is the total enthalpy in the
	free stream, $\Gamma_0$ is the absolute value of the vortex circulation,
	$l$ is the period of the vortex street along the $x$-axis,
	$1+U$ is the velocity of vortices in the reference frame fixed to
	the cylinder, and the bar signifies time averaging.
	
	The analysis of a time-averaged flow in
	\cite{AleksyukDirectnumericalsimulation2018} indicated that the
	main reason for the variation of $\overline{i_0}$ is the
	appearance of a negative correlation
	$\overline{\vec{u}'\cdot\nabla i'_0}$ induced by the pressure
	variation due to the nonlinear term in the energy conservation
	law. Here, $\vec{u}$ is the velocity vector,
	$\vec{u}'=\vec{u}-\overline{\vec{u}}$, and
	$i'_0=i_0-\overline{i_0}$.
	
	The present study focuses on the influence of the vortex wake
	structure on the total-enthalpy distribution. Developing the
	kinematical-explanation approach described above, we apply both
	direct numerical simulations and consider various equilibrium
	configurations of simplified point vortex models for several
	vortex street structures.
	
	We consider the problem of forced transverse oscillations of a
	circular cylinder in a uniform flow based on the direct numerical
	solution of Navier-Stokes equations by the finite-element method
	(Section~\ref{sec:NumericalSimulation}). This problem is known for
	a variety of flow regimes in the wake. The diagram of wake
	patterns can be found in 
	\cite{WilliamsonVortexformationwake1988,LeontiniWakestateenergy2006}. Among others it contains
	2S, P, 2P, and P+S modes. Here, the notation of each mode
	describes how the cylinder sheds vortices per one cycle of the
	oscillations (S and P mean a single vortex and a pair of
	vortices respectively). In Section~\ref{sec:Calculations}, we consider flows at
	$\Rey = 500$, $\M=0.4$, $\Pr = 0.72$ and different frequencies and
	amplitudes of forced oscillations to demonstrate the influence of
	the vortex street structure on the time-averaged total enthalpy
	distribution.
	
	In Section~\ref{sec:PVM}, several equilibrium configurations of
	point vortex models are considered to estimate the influence of
	vortex structure on the energy separation efficiency. The
	predictions based on these models are in good agreement with the
	computation results, given the limitations discussed in
	Section~\ref{sec:PVM}. We restrict our consideration to vortex
	streets combined of 2, 3, and 4 infinite chains of potential
	vortices. For the details of the properties of such models see
	\cite{KochinTheoreticalhydromechanics1964,MeleshkoDynamicsvortexstructures1993,Arefmotionthreepoint1996,BasuExploringdynamics2P2017,
		StremlerRelativeequilibriasingly2003}.
	
	\section{Problem formulation and numerical method} \label{sec:NumericalSimulation}
	In the inertial reference frame $(x, y)$, the circular cylinder
	oscillates transversely in a uniform flow. The transverse
	displacement of the cylinder is given by the expression
	$y_c(t)=A\sin(2\pi f t)$; here, $A$ and $f$ are amplitude and
	frequency. The fluid surrounding the cylinder is
	described by the model of a viscous perfect gas with constant
	specific heats, viscosity, and thermal conductivity.
	
	The problem is formulated in the Cartesian coordinate system
	$(x_1, x_2)$ fixed to the circular cylinder, with the origin
	located at its center (see Fig.~\ref{fig:PVM_Scheme}a).  The
	Navier--Stokes equations governing the compressible fluid flow are
	solved in primitive variables $\vec{Y}(\vec{x},t)=(p, u_1, u_2,
	T)^*$; here, $p$, $\vec{u}=(u_1, u_2)^*$ and $T$ are the
	dimensionless pressure, velocity vector, and temperature;
	$\vec{x}=(x_1, x_2)^*$; and $^*$ is transposition. After the
	transition from the conservative variables the  Navier--Stokes
	equations can be written in the following form
	\begin{equation}\label{eq:NS}
	\mat{A}_0 \vec{Y}_{,t} + \mat{A}_i \vec{Y}_{,i} = \left(\mat{K}_{ij}\vec{Y}_{,j}-\vec{P}_i\right)_{,i}+\vec{R}.
	\end{equation}
	The explicit expressions for matrices $\mat{A}_0$, $\mat{A}_i$,
	$\mat{K}_{ij}$ and vectors $\vec{P}_i$, $\vec{R}$ ($i, j = 1, 2$)
	are given in  \ref{appA}. The last vector on the right-hand side
	arises due to the use of the non-inertial reference frame. The
	repeated indices imply summation, and the short notation for the
	derivatives is used $(\cdot)_{,t}=\partial(\cdot)/\partial t$,
	$(\cdot)_{,1}=
	\partial(\cdot)/\partial x_1$, $(\cdot)_{,2}=\partial(\cdot)/\partial x_2$.
	
	All quantities are dimensionless; nondimensionalization is performed using the
	following formulas (here, dimensional quantities are denoted by
	primes)
	\begin{equation*}
	t=\frac{u_\infty t'}{d},\ \vec{x}=\frac{\vec{x}'}{d},\ p=\frac{p'}{\rho_\infty u_\infty^2},\ \vec{u}=\frac{\vec{u}'}{u_\infty},\  T=\frac{c_V T'}{u_\infty^2}.
	\end{equation*}
	The Reynolds $\Rey$, Prandtl $\Pran$, and Mach $\M$ numbers are
	defined by the following formulas:
	\begin{equation*}
	\Rey=\frac{\rho_\infty u_\infty d}{\mu},\
	\Pran=\frac{\mu c_p}{\kappa},\  \M=\frac{u_\infty}{c_\infty}.
	\end{equation*}
	Here, $d$ is the diameter of the cylinder; $\rho_\infty$, $p_\infty$, $u_\infty$ are the free-stream density, pressure, and velocity;
	$\kappa, \mu$ are the thermal conduction and viscosity coefficients; $c_V, c_p$ are the specific heats at constant volume and pressure;
	$c_\infty=\sqrt{{\gamma p_\infty}/ {\rho_\infty}}$ is the sonic velocity in the free stream; and $\gamma=1.4$ is the specific heat
	ratio. Amplitude $A$ and frequency $f$ of cylinder oscillations are also dimensionless: $A=A'/d$, $f=f'd/u_\infty$. In what follows, we
	will use the dimensionless parameter $F=f'/f'_0$ instead of $f$; here, $f'_0$ is the vortex shedding frequency for a fixed cylinder.

\begin{table}
	\begin{center}
		\begin{tabular}{l|llll}
			Domain  & $\Delta$ & $X_{in}$ & $X_{out}$ & $Y$ \\ [3pt]
			\hline
			Boundary layer & 0.0005 & - & - & - \\
			Near wake & 0.025 & 1.5 & 26 & 5 \\
			Middle wake & 0.05 & 3 & 40 & 8 \\
			Far wake & 0.25 & 6 & 80 & 16 \\
			Entire domain & 2.5 & 200 & 400 & 200 \\
		\end{tabular}
		\caption{Parameters of the mesh. The number of nodes is $N_{v}=930930$. The number of elements is $N_{e}=1861860$. $\Delta$ is the
			approximate size of triangular elements of the mesh, $X_{in}$, $X_{out}$, and $Y$ are the distances from the center of the cylinder
			to the inlet, outlet, and side boundaries of the corresponding subdomains.}{\label{tbl:Meshes}}
	\end{center}
\end{table}

	In the non-inertial reference frame, the following boundary
	conditions are assumed. On the cylinder surface, the velocity
	no-slip $\vec{u}=(0,0)$ and adiabatic-wall $\nabla
	T\cdot\vec{n}=0$ conditions are prescribed (here, $\vec{n}$ is a
	unit normal vector). At infinity, $\vec{u}=(1,-v_c)$,
	$p={1/{(\gamma\M^2)}}$, and $T={1/{[\gamma(\gamma-1)\M^2]}}$, where
	$v_c=y_{c,t}$.
	
	For numerical solution of this problem, we use the Galerkin
	least-squares (GLS) finite-element method on unstructured
	triangular meshes. The algorithms used have been successfully applied
	to the simulation of several problems of compressible and
	incompressible viscous flows around bodies (for details, see
	\cite{AleksyukDirectnumericalsimulation2018,AleksyukAnalysisthreedimensionaltransition2018,AleksyukFormationevolutiondecay2012}).

	\begin{figure}[!t]
		\centering
		\begin{subfigure}[b]{90mm} \centering
			\includegraphics[width=90mm] {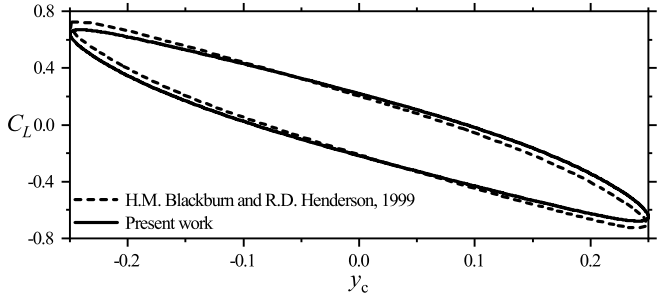}\vspace{5px}
		\end{subfigure}\\\vspace{5px}
		\caption{The dependence of lift coefficient $C_L$ on cylinder displacement $y_c$ for limit cycle at $F=0.89$, $A=0.25$, $\Rey = 500$. The dashed line is data from \cite{Blackburnstudytwodimensionalflow1999} and the solid line is the present results ($\M=0.1$ and $\Pr = 0.72$).}\label{fig:CLTest}
	\end{figure}

	Table~\ref{tbl:Meshes} shows the parameters of the mesh used,
	which is divided into several nested subdomains with different
	step sizes. The results of algorithm testing for similar regimes
	of the flow can be found in
	\cite{AleksyukDirectnumericalsimulation2018}. In this study, we
	enlarged the region of `Near wake' along and across the flow,
	since here we consider wider wakes due to cylinder oscillations
	and also because we are interested in the development of the wakes
	at longer distances downstream.
		
\begin{figure}[!t]
		\centering
		\begin{subfigure}[b]{90mm} \centering
			\includegraphics[width=90mm] {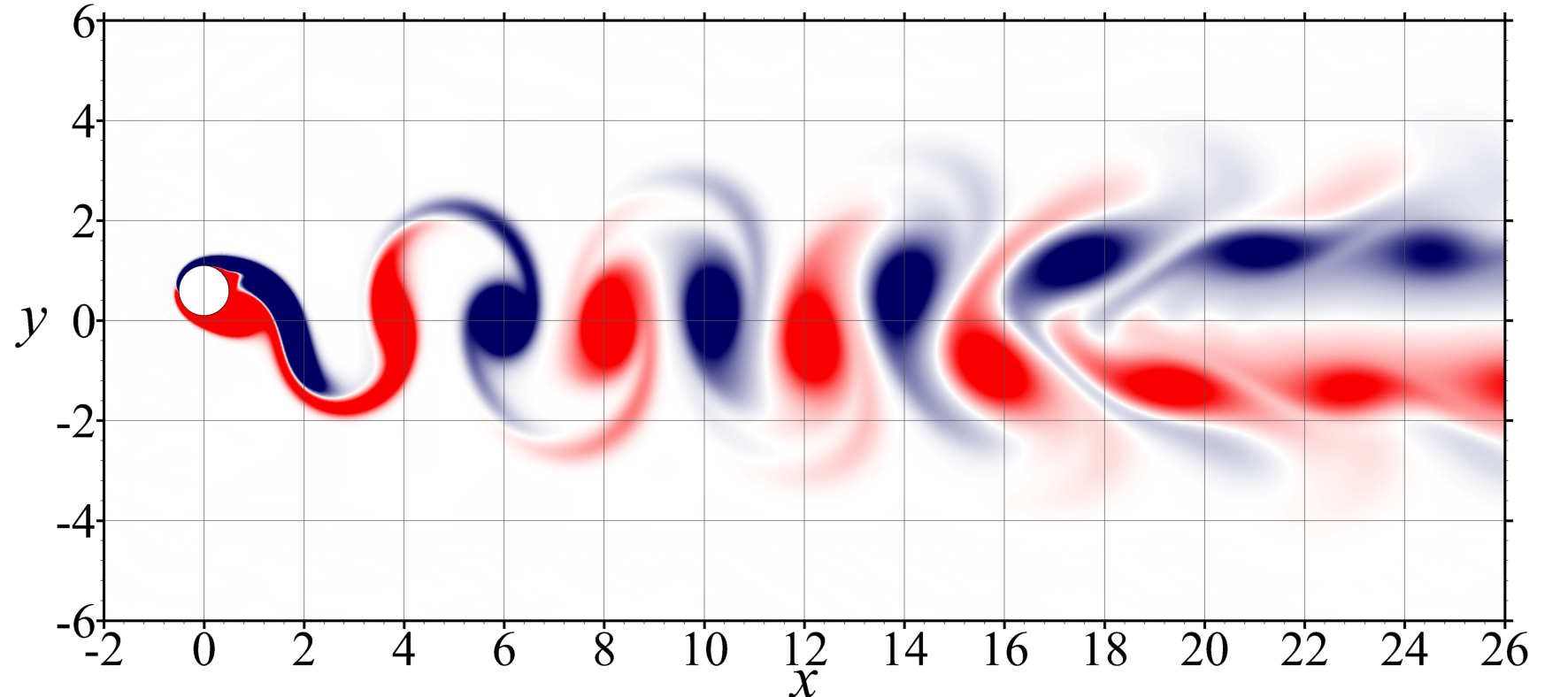}\vspace{5px}
			\caption{$A=0.6$}
		\end{subfigure}\\\vspace{5px}
		\begin{subfigure}[b]{90mm} \centering
			\includegraphics[width=90mm] {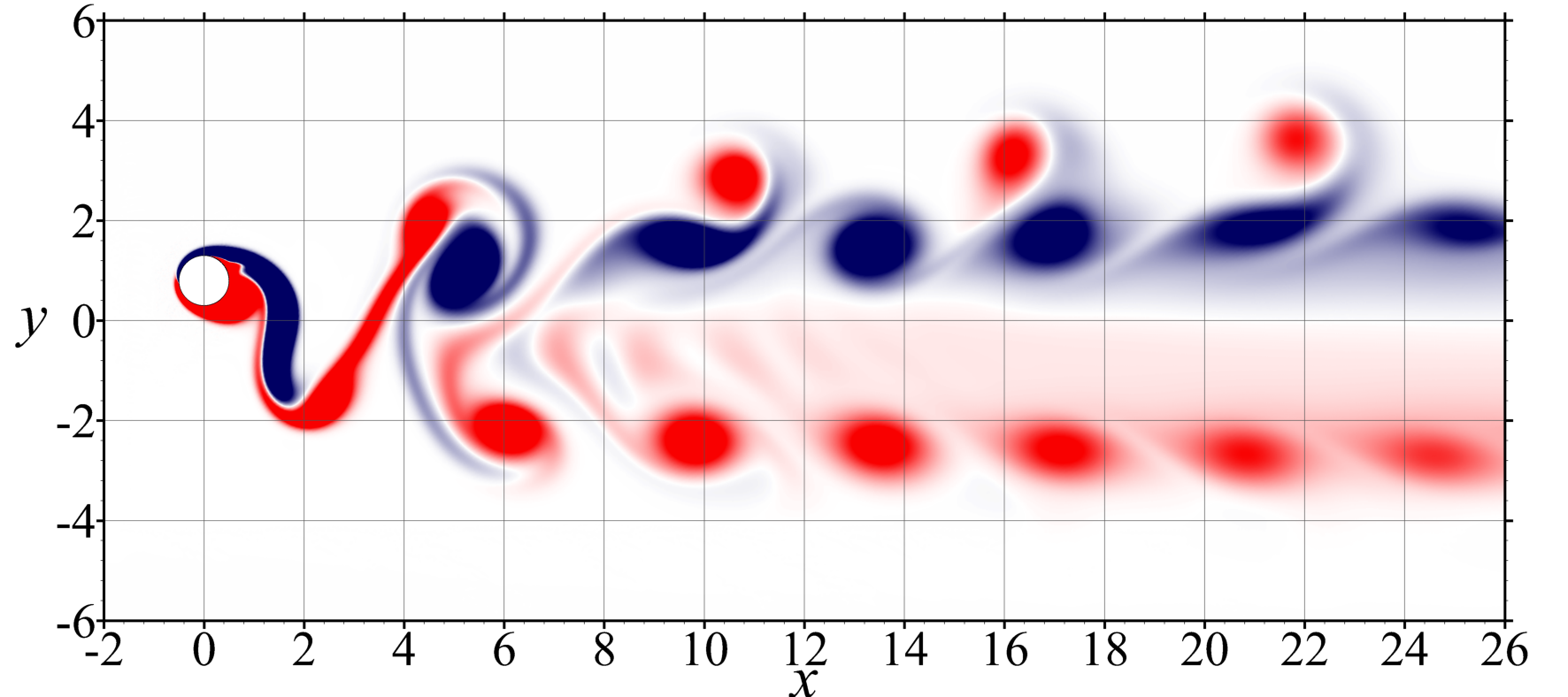}\vspace{5px}
			\caption{$A=0.8$}
		\end{subfigure}\\\vspace{5px}
			\begin{subfigure}[b]{90mm} \centering
			\includegraphics[width=90mm] {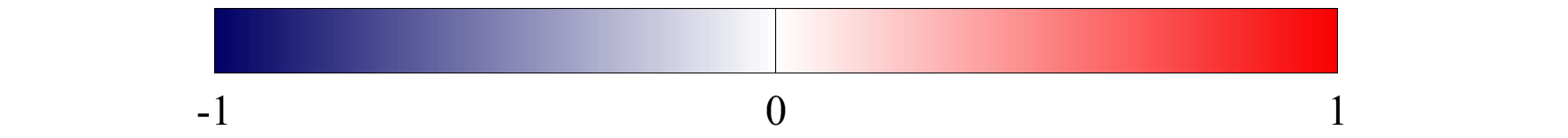}\vspace{5px}
			\caption{Color map for $\omega$}
		\end{subfigure}\\\vspace{5px}
		\caption{The transition from 2S (a) to P+S (b) shedding mode in the wake at $F=1.01$, $\Rey = 200$, $\M=0.1$ and $\Pr = 0.72$. Limit values on the color map are not the maximum and minimum of function $\omega$: function values greater than the upper limit (or less than the lower limit) are filled with one color corresponding to this limit. (For the interpretation of the references to color in this figure legend, the reader is referred to the web version of this article.)}\label{fig:PatternsTest}
	\end{figure}

	The numerical results in following section are restricted to $\Rey = 500$. This Reynolds number was chosen to reduce the viscous diffusion effect and to obtain periodic regimes of a two-dimensional compressible flow ($M=0.4$) around a fixed cylinder. In \cite{AleksyukDirectnumericalsimulation2018} (Figs.~10-12) it was shown that at $\Rey\ge500$ the effect of the Reynolds number is not as pronounced as at $\Rey<500$. It should be noted, that the real flow at $\Rey = 500$ is three-dimensional and turbulent. However, we believe that the described mathematical model takes into account enough underlying physics to qualitatively demonstrate the influence of a vortex street on the efficiency of energy separation (Section~\ref{sec:Calculations}) and to study the adequacy of simplified point vortex models predicting the efficiency of this process in a developed wake (Section~\ref{sec:PVM}). In \cite{Blackburnstudytwodimensionalflow1999} the incompressible two-dimensional flow past an oscillating cylinder at $\Rey=500$ and $A=0.25$ was studied. Figure~\ref{fig:CLTest} demonstrates the agreement in the lift coefficient $C_L(y_c)$ with these results at $\Rey=500$, $F=0.89$, $A=0.25$. Figure~\ref{fig:PatternsTest} shows the transition between two vortex shedding regimes at $\Rey=200$ and $F=1.01$: 2S mode at $A=0.6$ and P+S mode at $A=0.8$. These results are in agreement with the data in \cite{LeontiniWakestateenergy2006}, where it was shown that the boundary amplitude $A$ between two modes is approximately $0.7$ (see Fig.~5a and 7 in \cite{LeontiniWakestateenergy2006}).

	\section{Energy redistribution for different wake patterns}\label{sec:Results}
	Total enthalpy $i_0$ is considered in the coordinates $(x,
	y)=(x_1,x_2+y_c)$ ($u=u_1$, $v=u_2+v_c$): the cylinder has zero
	$x$-component of the velocity and performs only transverse
	oscillations. We define normalized total enthalpy $I_0$ by the
	expression
	\begin{equation}\label{eq:I0}
	I_0(x,y,t)=\frac{i_0-i_{0\infty}}{i_{0\infty}}, \ i_0(x,y,t)=\gamma T+0.5\left(u^2+v^2\right),
	\end{equation}
	and introduce the time-averaged flow parameters denoted with the bar: $\overline{f}(x,y)=(t_2-t_1)^{-1}\int_{t_1}^{t_2}f(x,y,t)dt$, for some function $f(x,y,t)$ and long enough time interval $t_2-t_1$, excluding initial stages of flow development. In the present work, time-averaged distributions were obtained by averaging over the intervals not less than $215$ (approximately 48 vortex shedding periods for a fixed cylinder) with the time step being equal to $0.2$. The minimum value of the normalized total enthalpy in the time-averaged flow is considered as the indicator of energy separation efficiency 
	\begin{equation*}
	E(\Omega)=\left|\min_{(x,y)\in\Omega}\overline{I_0}(x,y)\right|.
	\end{equation*}
	The value of $E$ depends on the flow subregion $\Omega$ under consideration, for example, it could be the vortex formation region or some parts of the developed wake.
	
	\begin{figure*}[!t]
		\centering
		\begin{subfigure}[b]{90mm} \centering
			\includegraphics[width=90mm] {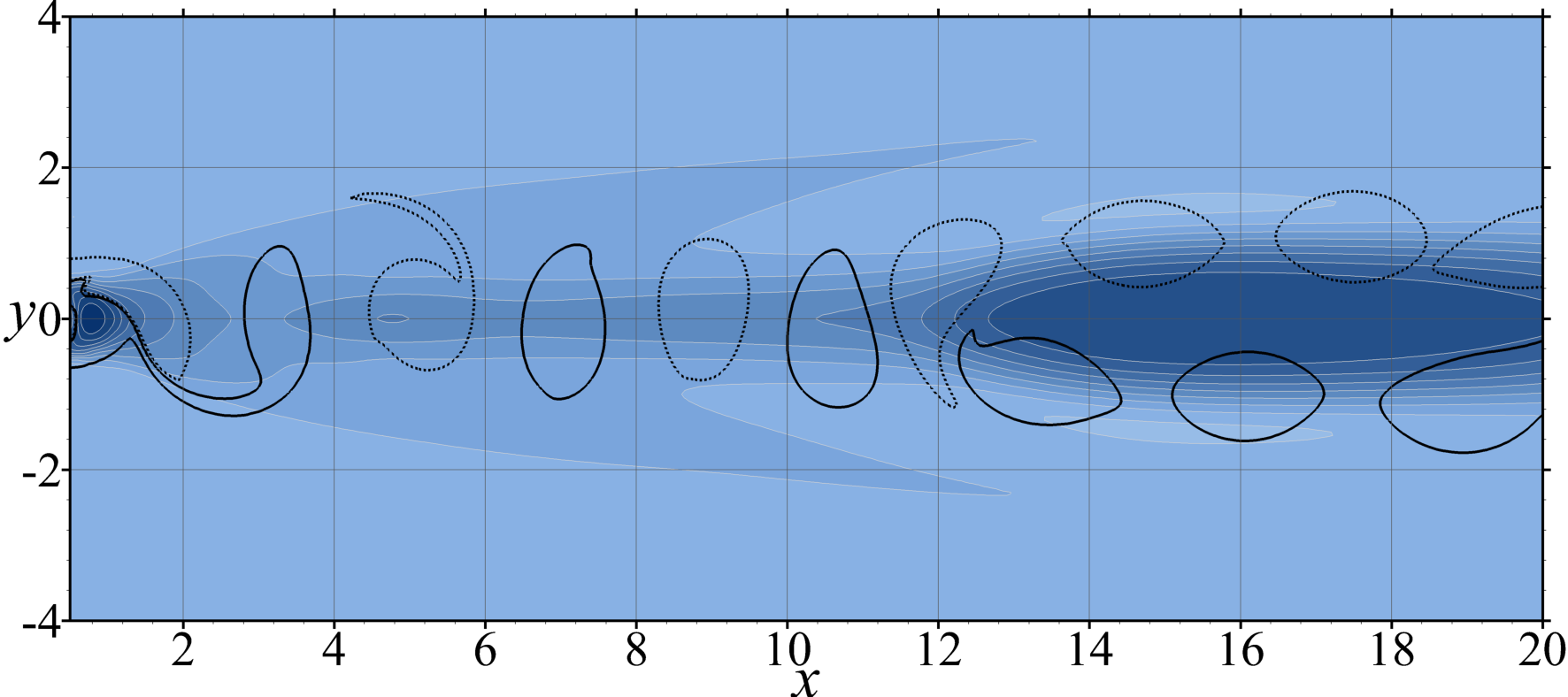}\vspace{5px}
			\caption{Fixed cylinder}
		\end{subfigure}~
		\begin{subfigure}[b]{90mm} \centering
			\includegraphics[width=90mm] {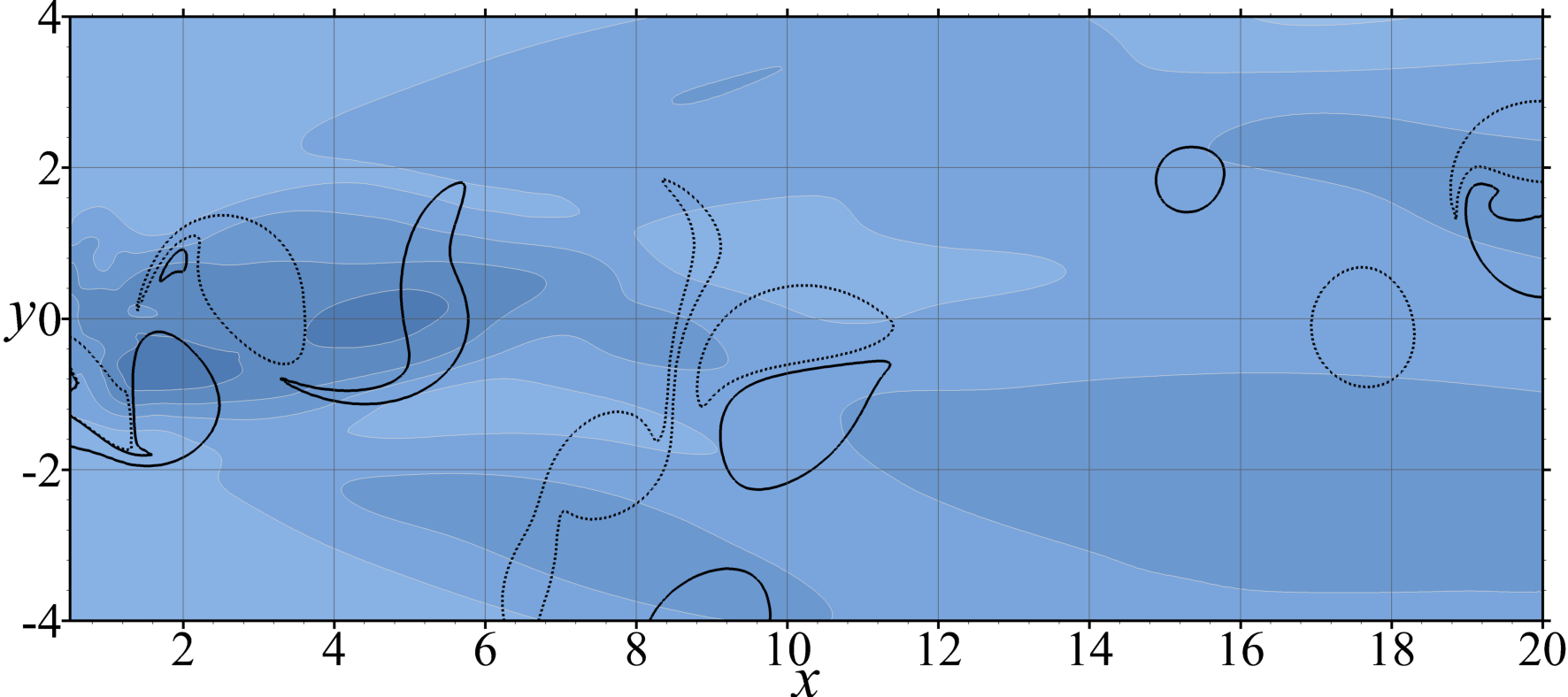}\vspace{5px}
			\caption{$A=1$, $F=0.5$}
		\end{subfigure}\\\vspace{5px}
		\begin{subfigure}[b]{90mm} \centering
			\includegraphics[width=90mm] {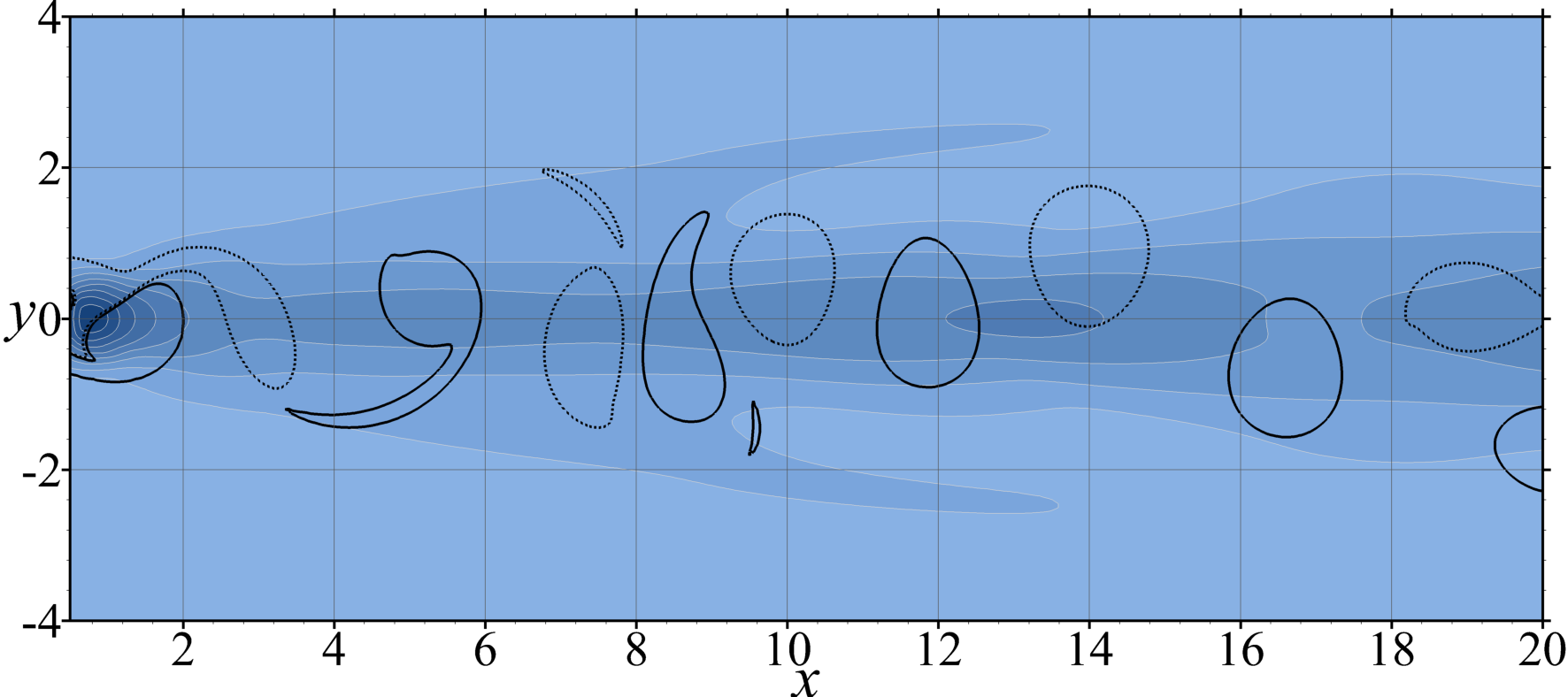}\vspace{5px}
			\caption{$A=0.25$, $F=0.7$}
		\end{subfigure}~
		\begin{subfigure}[b]{90mm} \centering
			\includegraphics[width=90mm] {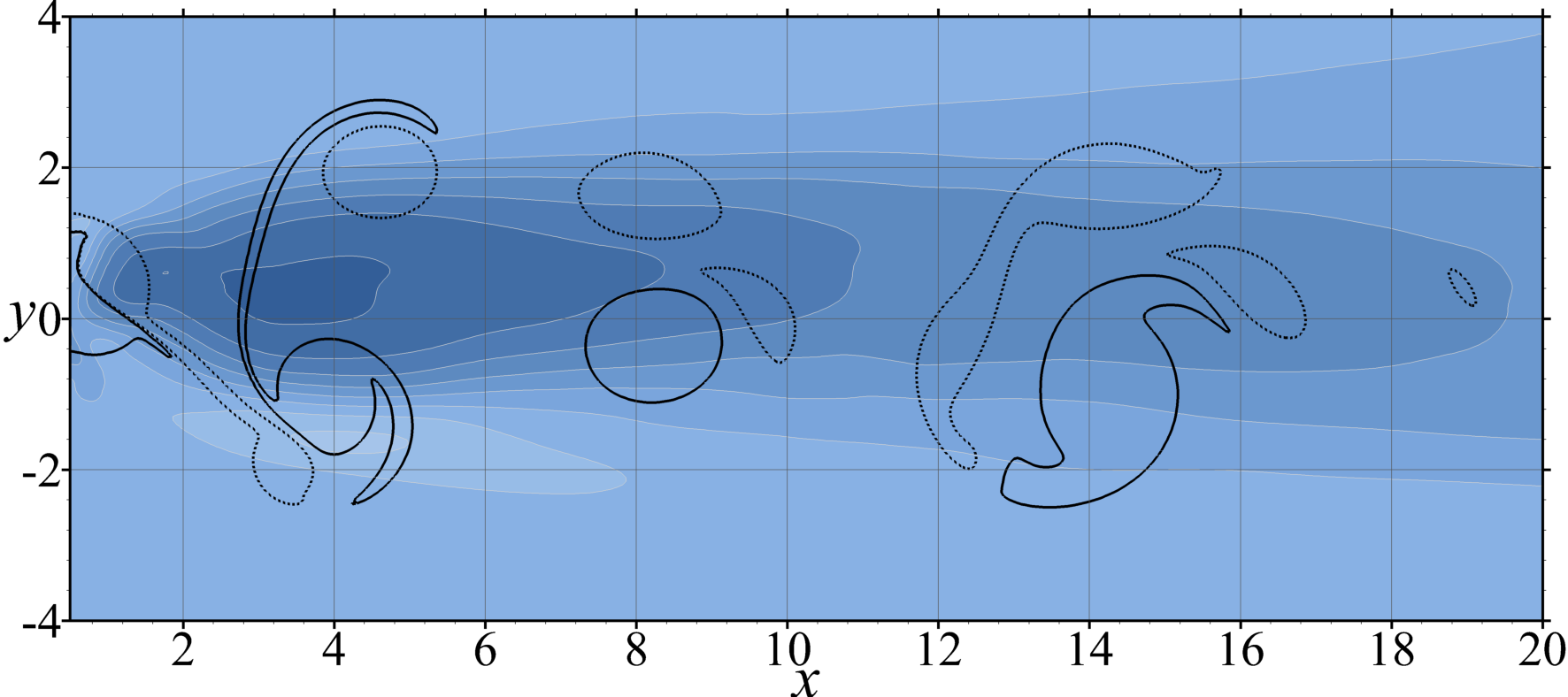}\vspace{5px}
			\caption{$A=0.75$, $F=0.7$}
		\end{subfigure}\\\vspace{5px}
		\begin{subfigure}[b]{90mm} \centering
			\includegraphics[width=90mm] {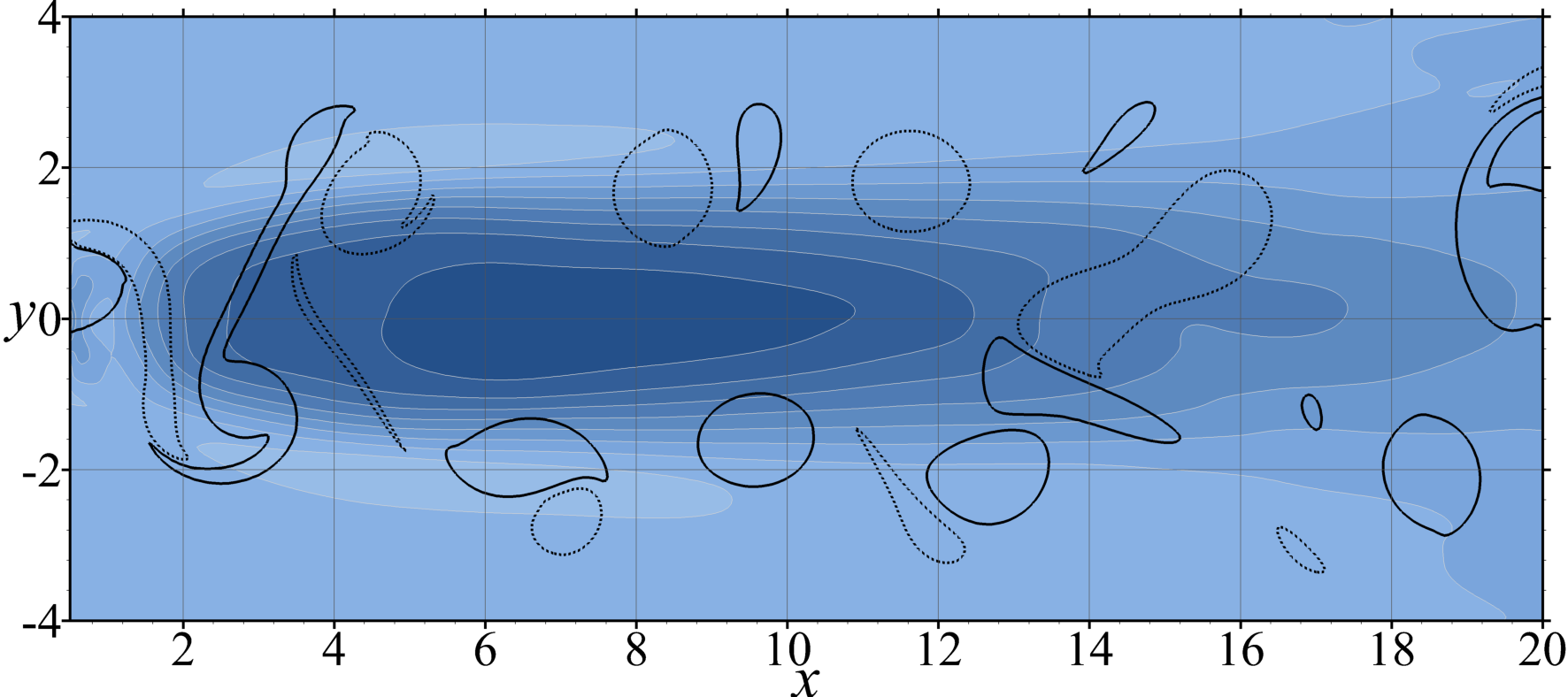}\vspace{5px}
			\caption{$A=0.5$, $F=1$}
		\end{subfigure}~
		\begin{subfigure}[b]{90mm} \centering
			\includegraphics[width=90mm] {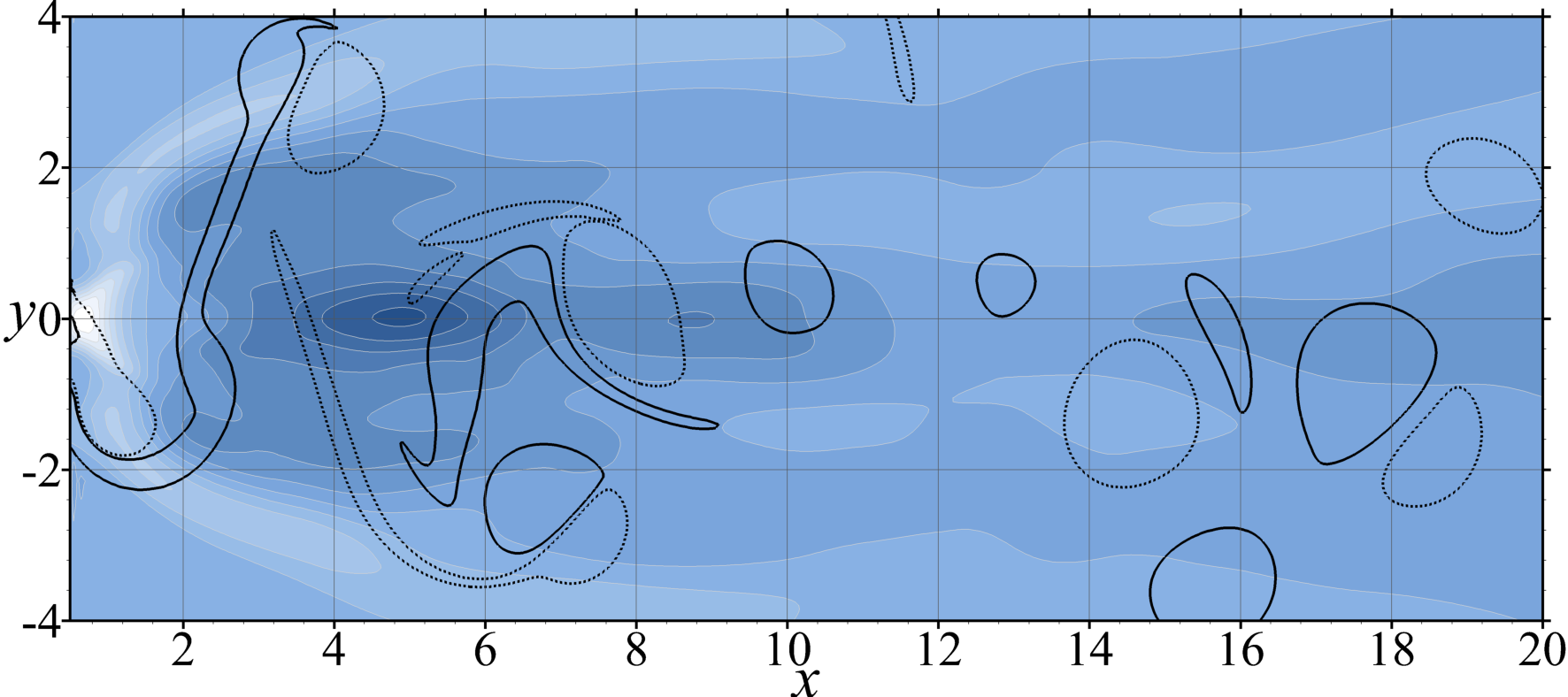}\vspace{5px}
			\caption{$A=1$, $F=1$}
		\end{subfigure}\\\vspace{5px}
		\begin{subfigure}[b]{90mm} \centering
			\includegraphics[width=90mm] {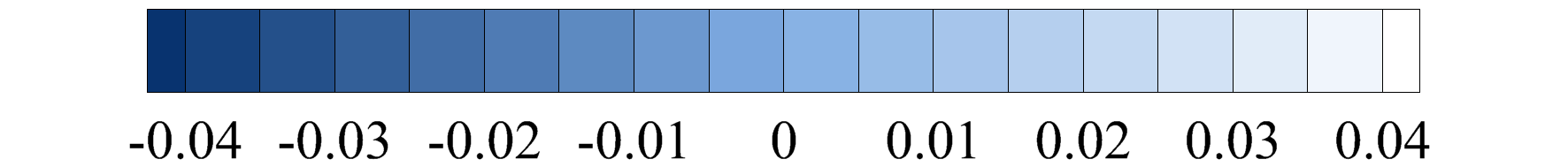}\vspace{5px}
			\caption{Color map for $\overline{I_0}$}
		\end{subfigure}\\
		\caption{The influence of the wake structure on the time-averaged total enthalpy distribution
			$\overline{I_0}$ at $\Rey = 500$, $\M=0.4$ and $\Pr = 0.72$. Here, solid and dashed lines correspond to positive and negative constant
			values of vorticity $\omega=\pm0.5$ at an arbitrary instant of time. (For the interpretation of the references to color in this figure legend, the
			reader is referred to the web version of this article.)}\label{fig:I0}
	\end{figure*}
	
	The equation for the total-enthalpy variation in the coordinates
	$(x, y)$ can be written as follows:
	\begin{equation}\label{eq:Mechanisms}
	\frac{D i_0}{Dt}=\underbrace{\frac{1}{\rho}\frac{\partial p}{\partial t}}_{\textstyle\mathcal{P}}+
	\underbrace{\frac{1}{\rho \Rey}\nabla\cdot\left(\mat{\tau}\cdot\vec{v}\right)}_{\textstyle\mathcal{A}}+
	\underbrace{\frac{1}{\rho\Rey\Pran}\nabla^2T}_{\textstyle\mathcal{Q}}.
	\end{equation}
	Here, $D/Dt=\partial/\partial t + \vec{v}\cdot\nabla$ is the
	material derivative and $\vec{v}=(u,v)^*$. The description of the
	contribution of each term in the equation for a fixed cylinder can
	be found in \cite{AleksyukDirectnumericalsimulation2018}. The reason for energy redistribution in the developed wake is related to
	the nonlinearity in the left-hand side of
	Eq.~\eqref{eq:Mechanisms} and to the action of term $\mathcal{P}$,
	whereas the other terms ($\mathcal{A}, \mathcal{Q}$) are
	negligible in the wake
	\cite{KurosakaEnergyseparationvortex1987,AleksyukDirectnumericalsimulation2018,EckertCrosstransportenergy1987}.
	The pressure fields in the wake are mainly determined by vortex
	dynamics. That is why we are interested in studying the effect of
	vortex street structures on the energy redistribution. This
	section demonstrates the influence of the wake pattern on the
	efficiency of energy separation based on both direct numerical
	simulation and simplified point vortex modeling of the wake.
	
	\subsection{Total-enthalpy distribution for a transversely oscillating cylinder}\label{sec:Calculations}
	Transverse oscillations of the cylinder cause energy transfer
	between the flow and the cylinder. For a period of time $t_1\le
	t\le t_2$, it can be expressed by the coefficient
	\begin{equation}\label{eq:CE}
	C_E=\int_{t_1}^{t_2}C_Lv_cdt.
	\end{equation}
	Equation ~\eqref{eq:CE}
	expresses the work done by the fluid. If $C_E<0$, energy transfers
	from the cylinder to the fluid, and vice versa. In the first case
	($C_E<0$), the shedding vortices tend  to suppress the cylinder
	oscillations, and in the second case ($C_E>0$) they tend to strengthen these
	oscillations. In this section, we consider the regimes with
	$C_E<0$.
	
	In Fig. ~\ref{fig:I0} we present the averaged distribution of total
	enthalpy for different frequencies $F$ and amplitudes $A$, and
	also the isolines $\omega=\pm\text{const}$ for a certain instant
	of time. The wake structure turns out to be mostly irregular;
	however, in most cases the time-averaged flow fields have reflective
	symmetry with respect to the wake centerline $y=0$. The loss of
	symmetry is clearly observed for  $A=1, F=0.5$ and $A=0.75, F=0.7$
	(Fig.~\ref{fig:I0}b,~d). At $A=0.75$ and $F=0.7$
	(Fig.~\ref{fig:I0}d), the vortices are shed in accordance with the
	P+S scheme: a single vortex on one side and a pair of vortices on
	the other. Therefore, as a result of averaging, the region of
	reduced total enthalpy is shifted away from the centerline. At
	$A=1$ and $F=0.5$ (Fig.~\ref{fig:I0}b), the widest region of
	lowered $\overline{I_0}$ is observed among the considered regimes.
	The vortex dynamics is complicated and it is difficult to
	construct the pattern in this case. Most of the time, the vortices
	are shed in pairs, but their trajectories are complex and differ
	from one cycle to another.
	
	From Fig.~\ref{fig:I0}, it is clear that when the vortices of
	opposite signs move with lesser transverse distance, smaller
	values of $E$ are attained (in the general case, it is not true, see
	Section~\ref{sec:PVM}). For example, see Figs.~\ref{fig:I0}a,~c,~e
	for $x\lesssim12$ and $x\gtrsim12$; and  Figs.~\ref{fig:I0}d,~f
	for $x\lesssim6$ and $x\gtrsim6$. Another finding is the
	following: if the positive vortex moves above (at a greater $y$ value than)
	the negative one, then the value of $\overline{I_0}$ between them
	is positive, and vice versa. Such thin bands of positive
	$\overline{I_0}$ are clear in Figs.~\ref{fig:I0}d,~e,~f for the
	regions of motion of some vortex pairs. These features can be
	explained using the simplified point vortex models; see
	Section~\ref{sec:PVM}.

	Particular attention should be paid to the effect shown in
	Fig.~\ref{fig:I0}a: after the change in the wake structure  the
	minimum value of the total enthalpy ($E\approx 0.034$) becomes
	closer to its values in the formation region ($E\approx 0.044$),
	while the region of reduced values is much larger. This effect is
	attributable to changing the vortex structure. In the next
	section, it  will be considered in more detail.
	
\begin{figure}[!t]
		\centering
		\begin{subfigure}[b]{90mm} \centering
			\includegraphics[width=90mm] {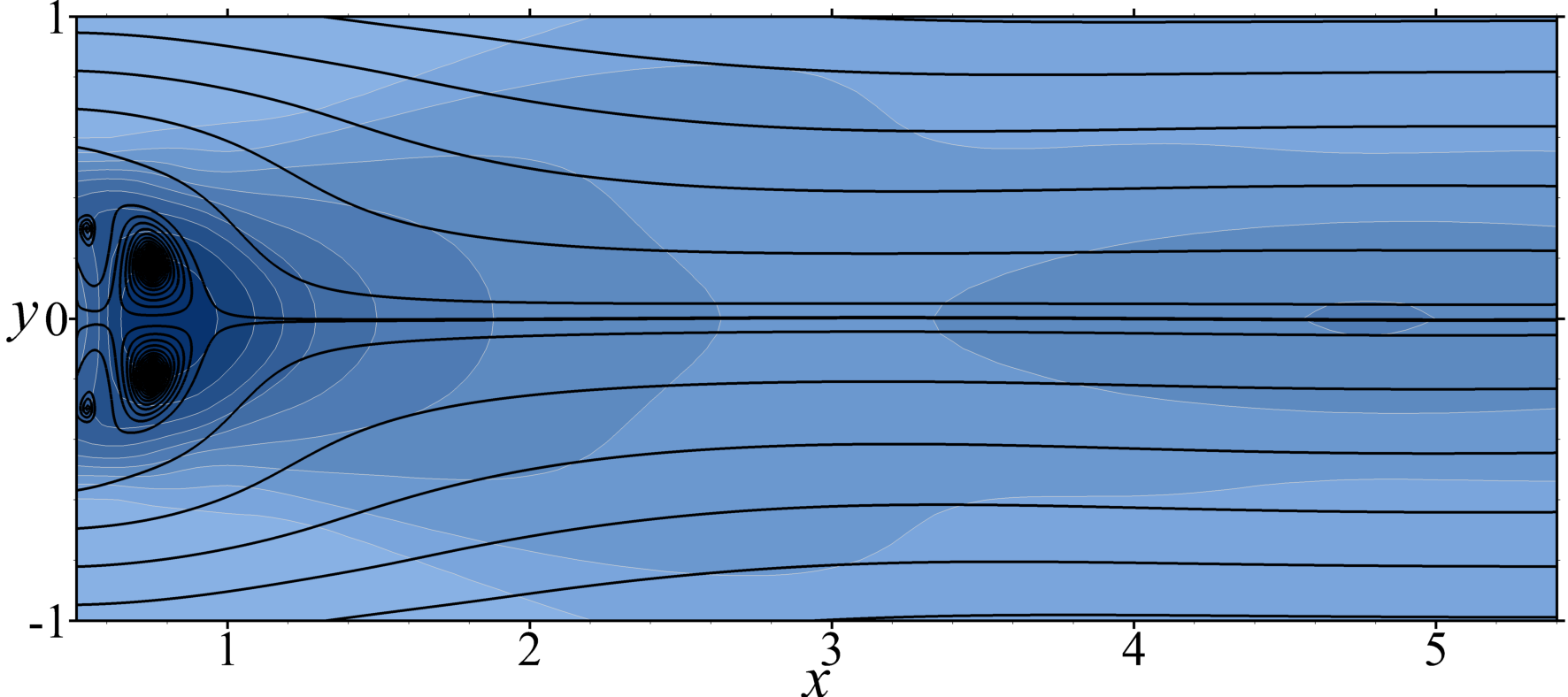}\vspace{5px}
			\caption{Fixed cylinder}
		\end{subfigure}\\\vspace{5px}
		\begin{subfigure}[b]{90mm} \centering
			\includegraphics[width=90mm] {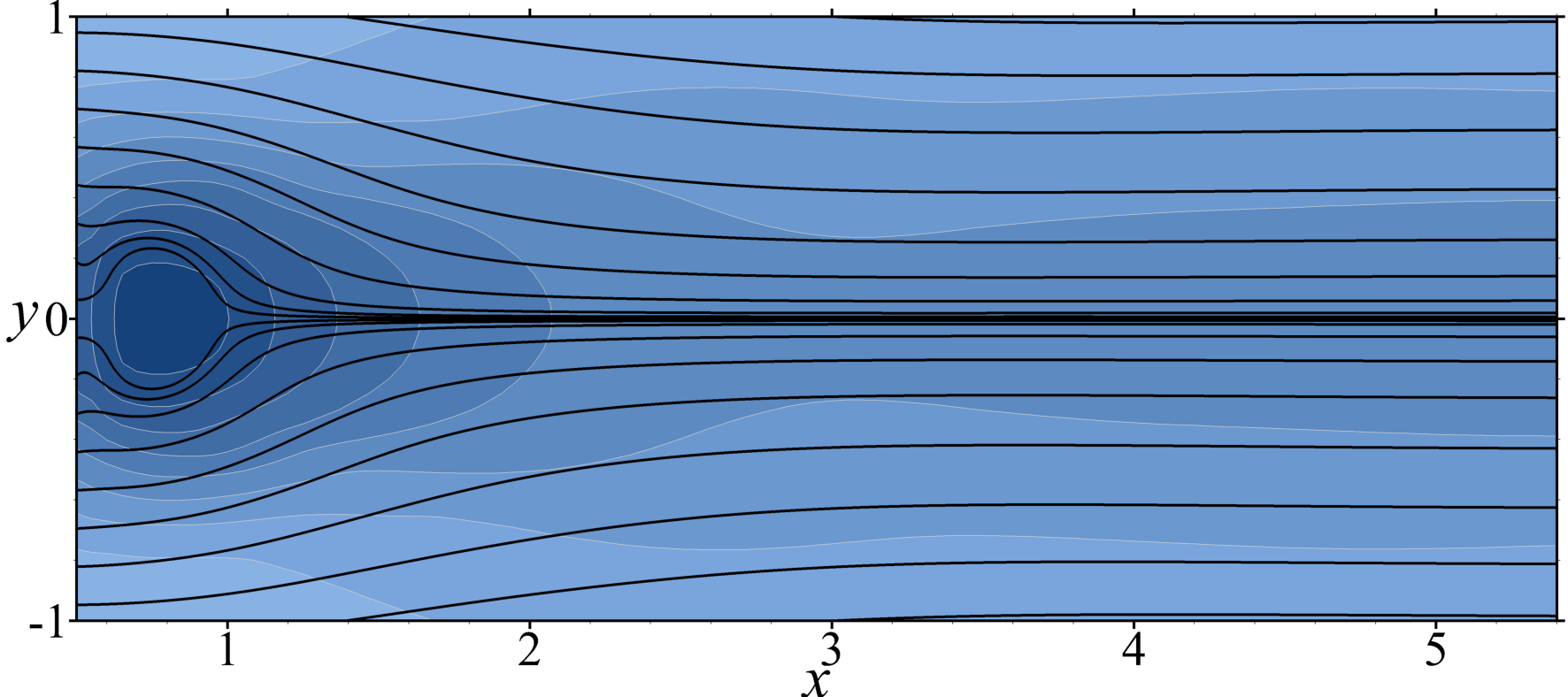}\vspace{5px}
			\caption{$A=0.25$, $F=0.7$}
		\end{subfigure}\\\vspace{5px}
		\begin{subfigure}[b]{90mm} \centering
			\includegraphics[width=90mm] {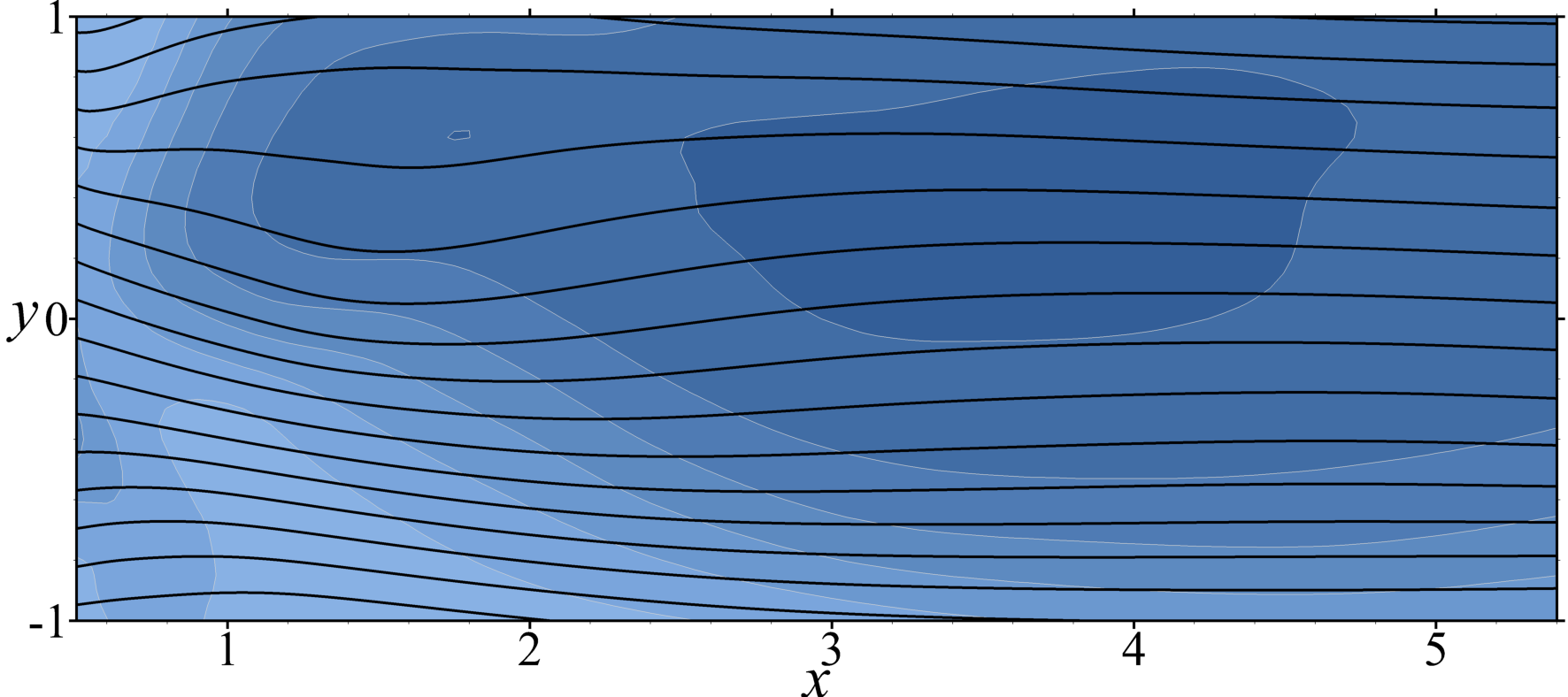}\vspace{5px}
			\caption{$A=0.75$, $F=0.7$}
		\end{subfigure}\\\vspace{5px}
		\begin{subfigure}[b]{90mm} \centering
			\includegraphics[width=90mm] {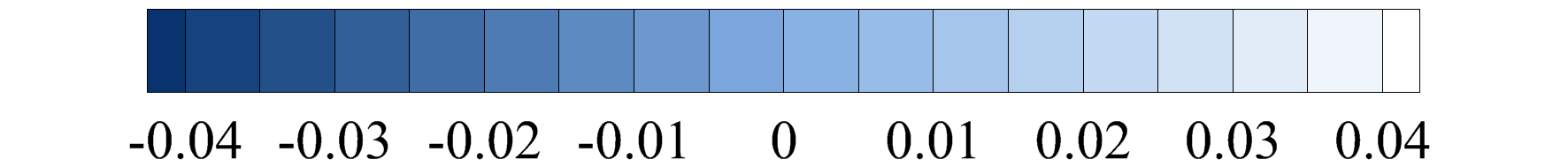}\vspace{5px}
			\caption{Color map for $\overline{I_0}$}
		\end{subfigure}\\
		\caption{Time-averaged total enthalpy distribution $\overline{I_0}$ and streamlines
			at $\Rey = 500$, $\M=0.4$ and $\Pr = 0.72$. (For the interpretation of the references to color in this figure legend, the reader is
			referred to the web version of this article.)}\label{fig:Streamlines}
	\end{figure}

	As the amplitude of oscillations becomes high enough, the
	formation region is no longer the region with minimal values of
	$\overline{I_0}$. Moreover, in this region much greater values of
	$\overline{I_0}$ can be attained ($\max\overline{I_0}\approx
	0.045$ for $A=1$ and $F=1$, see Fig.~\ref{fig:I0}f). For a fixed
	cylinder, on the contrary, the smallest values are
	usually observed in this region. As shown numerically for a fixed cylinder
	\cite{AleksyukDirectnumericalsimulation2018}, the region
	responsible for the decrease in $\overline{I_0}$ in the vortex
	formation zone does not affect the decrease of $\overline{I_0}$ in
	the developed vortex street. It is related to the formation of
	recirculation zones near the body in the time-averaged flow fields,
	where the lowest values of $\overline{I_0}$ are observed; see, for
	example Fig.~\ref{fig:Streamlines}a. If the oscillation amplitude
	$A$ is not high, there are still some recirculation zones in the
	time-averaged flow in which one can find the minimum of
	$\overline{I_0}$ (Fig.~\ref{fig:Streamlines}b). However, when $A$
	is high enough the time-averaged recirculation zones disappear
	(Fig.~\ref{fig:Streamlines}c) and the mechanisms
	\cite{AleksyukDirectnumericalsimulation2018} of reduction of
	$\overline{I_0}$  in the formation region are absent. Thus, the
	minimal values of $\overline{I_0}$ are not observed in the
	formation region in Figs.~\ref{fig:I0}b,~d-f and
	Fig.~\ref{fig:Streamlines}c. Another consequence of this is that
	the distribution of $\overline{I_0}$ in the formation region has
	a greater impact on $\overline{I_0}$ in the developed
	wake (in terms of analysis of the time-averaged mechanisms
	\cite{AleksyukDirectnumericalsimulation2018}).
	
	Among the regimes considered in Fig.~\ref{fig:I0}, in the
	developed wake the value of $E$ has its maximum for a fixed
	cylinder, $E\approx0.035$. With the increase in amplitude $A$ at
	$F=1$, the efficiency decreases ($E\approx 0.033$ at $A=0.5$ and
	$E\approx 0.031$ at $A=1$), but for $A=0.5$ the area of the region
	of reduced
	$\overline{I_0}$ is maximal. The same behavior can be observed for
	$F=0.7$ ($E\approx0.039$ at $A=0.25$ and $E\approx0.026$ at
	$A=0.75$). It should be noted that the data presented are not
	sufficient to make definitive conclusions on the influence of $A$
	or $F$.
	
	\subsection{Estimates of energy separation efficiency based on the point vortex models}\label{sec:PVM}
	
		\begin{figure*}[!t]
		\centering
		\begin{subfigure}[b]{60mm} \centering
			\includegraphics[width=60mm] {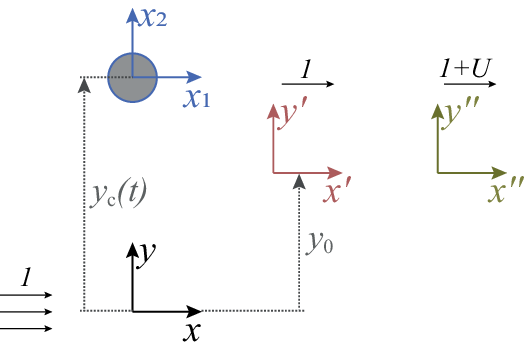}\vspace{5px}
			\caption{Coordinate systems}
		\end{subfigure}~
		\begin{subfigure}[b]{40mm} \centering
			\includegraphics[width=40mm] {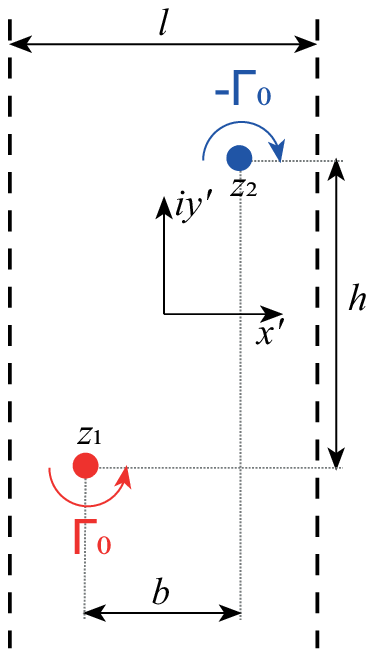}\vspace{5px}
			\caption{N=2}
		\end{subfigure}~
		\begin{subfigure}[b]{40mm} \centering
			\includegraphics[width=40mm] {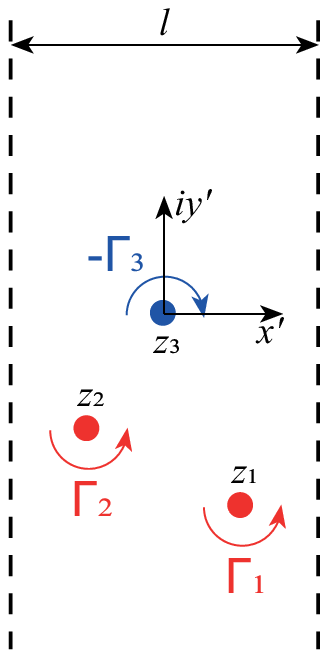}\vspace{5px}
			\caption{N=3}
		\end{subfigure}~
		\begin{subfigure}[b]{40mm} \centering
			\includegraphics[width=40mm] {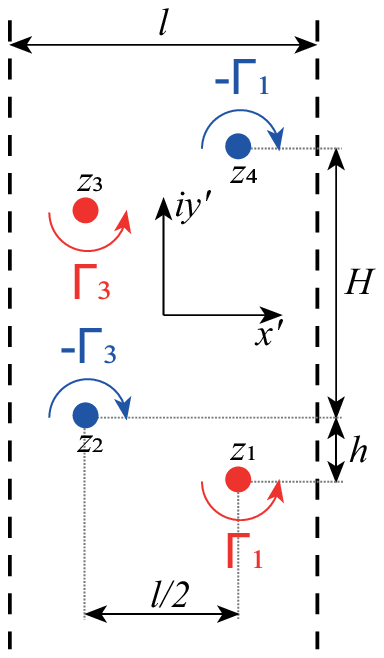}\vspace{5px}
			\caption{N=4}
		\end{subfigure}\\
		\caption{Schemes for different coordinate systems (a) and for one period of the considered point vortex models with $N=2$, $3$ and $4$ (b-d).} \label{fig:PVM_Scheme}
	\end{figure*}

	Since the main mechanism of energy separation in the developed
	wake is the time variation of pressure, it is possible to estimate
	the effect of the vortex wake structure on the energy separation
	efficiency using a simplified model of a vortex
	wake formed by potential vortices. A similar analysis was carried
	out earlier for the classical staggered K\'arm\'an street
	\cite{KurosakaEnergyseparationvortex1987}. To be complete, we
	repeat the assumptions required for calculating $\overline{I_0}$.

	The flow is assumed to be inviscid and non-heat-conducting. It is further assumed that the wake has a longitudinal period $l$; vortices in the wake move with constant
	velocity $1+U$ along the $x$-axis (velocity at infinity is 1 in
	dimensionless variables), which is greater than zero and less than
	1, so $-1<U<0$. We introduce the following coordinate systems
	(Fig.~\ref{fig:PVM_Scheme}a).
	\begin{itemize}
		\item[O1:] The flow at infinity is at rest, i.e. 
		
		$x'=x-t$, $y'=y-y_0$, where $y_0$ is constant; \\
		$u'(x',y',t)=u(x'+t,y'+y_0,t)-1$,\\ $v'(x',y',t)=v(x'+t,y'+y_0,t)$.
		\item[O2:] The flow in the wake is stationary, i.e. 
		
		$x''=x-(U+1)t$, $y''=y-y_0$; \\
		$u''(x'',y'')=u(x''+Ut+t,y''+y_0,t)-(U+1)$,\\ $v''(x'',y'')=v(x''+Ut+t,y''+y_0,t)$.
	\end{itemize}
	
	In O2 the flow is stationary, so the total enthalpy $i''_0=\gamma T+0.5(u''^2+v''^2)$ is constant along streamlines; we assume that this constant is the same for any streamline (for example, for isentropic irrotational flow). Thus, using the value of $i''_0$ at infinity,
	\begin{equation}\label{eq:PVM_I0_2}
	i''_0(x'', y'')=\gamma T+0.5(u''^2+v''^2)=\gamma T_\infty+0.5U^2.
	\end{equation}
	Here, $T_\infty$ is the dimensionless temperature at infinity, and $\gamma T_\infty=i_{0\infty}-0.5$. Excluding temperature $T$ from Eqs.~\eqref{eq:I0} and \eqref{eq:PVM_I0_2}, one can obtain the following expression.
	\begin{equation*}
	i_0(x,y,t)-i_{0\infty}=(1+U)(u-1).
	\end{equation*}
	For theoretical estimations in the present section purely periodic flows are considered, hence, in the definition of time-averaging (given in the beginning of Section~\ref{sec:Results}) $t_2-t_1$ equals one time period. Moreover, time-averaging coincides with spatial averaging along the $x$ axis and
	\begin{equation*}
	\overline{u}(y)=\frac{1}{t_2-t_1}\int\limits_{t_1}^{t_2}u(x,y,t)dt=U+1+\frac{1}{l}\int\limits_{\xi}^{\xi+l}u''(\xi,y-y_0)d\xi,
	\end{equation*}
	\begin{equation*}
	\overline{u'}(y')=\frac{1}{t'_2-t'_1}\int\limits_{t'_1}^{t'_2}u'(x',y',t)dt=U+\frac{1}{l}\int\limits_{\xi}^{\xi+l}u''(\xi,y')d\xi.
	\end{equation*}
	Here, $(U+1)(t_2-t_1)=U(t'_2-t'_1)=l$.
	Hence,
	\begin{equation}\label{eq:PVM_I0_General}
	\overline{I_0}(y)=\frac{(1+U)}{i_{0\infty}}\overline{u'}(y-y_0).
	\end{equation}
	The efficiency of energy separation equals $E=(1+U)|\overline{u'}_{\min}|/i_{0\infty}$, where $\overline{u'}_{\min}=\min_{y'}\overline{u'}(y')$.

	Thus, to estimate the energy separation efficiency, it is
	necessary to construct the velocity fields for different
	configurations of a periodic vortex wake. For this purpose, we use
	the point vortex models of the wake, which were studied previously
	in numerous works; see, for example,
	\cite{KochinTheoreticalhydromechanics1964,MeleshkoDynamicsvortexstructures1993,Arefmotionthreepoint1996,
		BasuExploringdynamics2P2017,StremlerRelativeequilibriasingly2003}.
	
	It is worth giving a few comments on the applicability of
	these models. The point vortex models do not reflect the shape of
	the vortices and predict incorrect velocity distribution in the
	vortex cores (additional remarks on it will be made in the next subsection). In addition, the considered model does not
	reproduce a variation in the mutual arrangement of the vortices.
	Partially it could be taken into account by considering
	non-equilibrium regimes; see, for example,
	\cite{Arefmotionthreepoint1996}. Not all of the considered
	configurations are stable and can be realized in nature
	\cite{KochinTheoreticalhydromechanics1964,Arefmotionthreepoint1996,BasuExploringdynamics2P2017}.
	Moreover, this model assumes that the fluid is incompressible, and
	the compressibility effects could change some parameters of the
	flow. For example, in \cite{CrowdySpeedKarmanpoint2017} one can
	find theoretical estimates of variations in $U$ for a weakly
	compressible fluid. Finally, viscous diffusion is absent in the
	point-vortex model. Nevertheless, these models provide a very
	convenient tool for understanding the general ways of increasing/decreasing the energy separation effect by changing the
	intensity and mutual arrangements of vortices.

	In a general case, let there be a periodic vortex wake with a
	longitudinal period $l$, which is formed by $N$ infinite chains of
	vortices, see Fig.~\ref{fig:PVM_Scheme}b-d. In each chain, we
	denote the complex-valued coordinates of the vortices defining the
	mutual position of the chains by $z_k=x'_k+iy'_k$, $k=1, ..., N$.
	The velocity field created by a system of point vortices is given
	by
	\begin{equation*}
	u'-iv'=\sum_{k=1}^{N}\frac{\Gamma_k}{2li}\cot\frac{\pi}{l}(z-z_k).
	\end{equation*}
	Using this equation, one can obtain the expression for
	$\overline{u'}$:
	\begin{equation}\label{eq:PVM_u'}
	\overline{u'}=-\sum_{k=1}^{N}\frac{\Gamma_k}{2l}\sign(y'-y'_k).
	\end{equation}
	
	The velocity of each vortex $z_\alpha$ is determined from the
	relation
	\begin{equation}\label{eq:PVM_VortexVelocity}
	u'_\alpha-iv'_\alpha=\sum_{k=1, k\neq\alpha}^{N}\frac{\Gamma_k}{2li}\cot\frac{\pi}{l}(z_\alpha-z_k).
	\end{equation}
	For equilibrium solutions, the positions of the vortices relative
	to each other do not change, so
	\begin{equation}\label{eq:PVM_Equilibrium}
	u'_\alpha=U,\quad v'_\alpha=0,\quad \sum_{k=1}^{N}\Gamma_k=0.
	\end{equation}
	Here, we use the fact that the motion occurs along the $x$-axis
	with velocity $U<0$. The last relation in
	Eqs.~\eqref{eq:PVM_Equilibrium} is obtained from the first two
	conditions after multiplying by $\Gamma_\alpha$
	Eq.~\eqref{eq:PVM_VortexVelocity} and summing over index $\alpha$.

	We introduce  the following parameters, which
	characterize the flow: $\beta=lU/\Gamma<0$ and
	$\alpha_k=\Gamma_k/\Gamma$,
	$\Gamma=|\Gamma_1|+|\Gamma_2|+...+|\Gamma_N|$,
	$\alpha_1+\alpha_2+...+\alpha_{N}=0$, $k=1 ,..., N$. If $\alpha_k$ and $\beta$ are given, one can obtain a solution (or solutions) of Eqs.~\eqref{eq:PVM_VortexVelocity} and \eqref{eq:PVM_Equilibrium}, which defines certain mutual arrangement of vortices up to scaling $l$. From Eqs.~\eqref{eq:PVM_I0_General} and \eqref{eq:PVM_u'} we have
	\begin{equation*}
	\overline{I_0}=-\frac{\Gamma}{2li_{0\infty}}(1+\beta\frac{\Gamma}{l})
	\sum_{k=1,~k\neq m}^{N}\alpha_k[\sign(y'-y'_k)-\sign(y'-y'_m)]
	\end{equation*}
	for any $m=1, 2,..., N$.  For
	any fixed mutual arrangement of vortices up to scaling $l$, with given $\alpha_k$ and $\beta$, the maximum of $E$ (the minimum of $\overline{I_0}$) is attained for
	$\beta\Gamma/l=-0.5$ (since positive factor $(\Gamma/l)(1+\beta\Gamma/l)$ is maximal), i.e.
	\begin{equation*}
	E=\frac{\Gamma}{4l i_{0\infty}}
	\max_{y'}\left\{\sum_{k=1,~k\neq m}^{N}\alpha_k[\sign(y'-y'_k)-\sign(y'-y'_m)]\right\}.
	\end{equation*}
	Since $\sum_{k=1}^{N}|\alpha_k|=1$ and the absolute value of the
	expression in square brackets is less than $2$, the following
	estimate is valid
	\begin{equation}\label{eq:Emax}
	E\le\frac{\Gamma}{2l i_{0\infty}}\min_k(1-|\alpha_k|)\le\frac{\Gamma(N-1)}{2l i_{0\infty}N}=-\frac{N-1}{4\beta i_{0\infty}N}.
	\end{equation}
	Furthermore, since $i_{0\infty}>0.5$ (in our calculations $M=0.4$
	and $i_{0\infty}=129/8$), $E\le\Gamma(N-1)/(lN)$.
	
	We  can use Rankine vortices instead of potential ones to better
	approximate the velocity field in the vortex cores. Despite the fact that the flow inside vortex cores is no longer irrotational, we still assume that Eq.~\eqref{eq:PVM_I0_General} is approximately valid. It is true if $\int_{\xi}^{\xi+l}i''_0(\xi,y')d\xi/l\approx i''_{0\infty}$ (for example, if $i''_0(x',y')=i''_{0\infty}$, as was assumed above). Thus, one should modify only the expression for $\overline{u'}$, instead of Eq.~\eqref{eq:PVM_u'},
	\begin{align}\label{eq:PVM_u'_RV}
	\begin{split}
	\overline{u'}=&\sum_{k=1}^{N}\frac{\Gamma_k}{2l}\sign(y'-y'_k)\left[\theta_k(y')\frac{2}{\pi}\arctan\sqrt{\left(\frac{R_k}{y'-y'_k}\right)^2-1}-1\right]\\
	&-\sum_{k=1}^{N}\frac{\Gamma_k}{\pi l}\theta_k(y')\frac{y'-y'_k}{R_k}\sqrt{1-\left(\frac{y'-y'_k}{R_k}\right)^2}.
	\end{split}
	\end{align}
	Here, $R_k$ is the radius of the core of vortices in $k$-chain and
	$\theta_k(y')$ equals 1 if $(y'-y'_k)^2<R_k^2$, otherwise it is
	zero. This function is continuous, unlike the case of potential
	vortices for which it is piece-wise constant.
	
	Below, we consider three cases shown in Fig.~\ref{fig:PVM_Scheme}b-d:
	classical symmetric and staggered K\'arm\'an streets with $N=2$,
	equilibrium configurations of three vortices with $N=3$, and
	a symmetric wake with $N=4$.
	
	\subsubsection{N=2}\label{sec:N=2} Let us introduce additional
	notations for this subsection (Fig.~\ref{fig:PVM_Scheme}b):
	$\Gamma_1=-\Gamma_2=\Gamma_0=0.5\Gamma>0$ and $z_2-z_1=b+ih$,
	where $b, h\ge0$. For symmetric ($b=0$) and staggered ($b=0.5l$)
	K\'arm\'an streets, one can obtain
	\begin{align}\label{eq:E_N=2}
	\begin{split}
	E=
	\begin{cases}\frac{\Gamma}{2li_{0\infty}}\left(1-\frac{\Gamma}{4l}\coth\frac{\pi h}{l}\right), \quad b=0,\\
	\frac{\Gamma}{2li_{0\infty}}\left(1-\frac{\Gamma}{4l}\tanh\frac{\pi h}{l}\right), \quad b=0.5l,\end{cases}
	\end{split}
	\end{align}
	due to the fact that $\overline{u'}_{min}=-\Gamma_0/l$ at $y'_1<y'<y'_2$ and $U=-(\Gamma_0/2l)\coth(\pi h/l)$ or $U=-(\Gamma_0/2l)\tanh(\pi h/l)$ for
	a symmetric or a staggered K\'arm\'an street. In terms of the previous analysis, $\alpha_1=-\alpha_2=0.5$
	and $\beta=-0.25\coth(\pi h/l)$ or $\beta=-0.25\tanh(\pi h/l)$ for a symmetric or a staggered K\'arm\'an street.
	
	Let us now consider the effect of parameters in
	Eq.~\eqref{eq:E_N=2} on the energy separation efficiency. When
	other parameters are fixed, the optimal value of $\Gamma/l$ is
	$2\tanh(\pi h/l)$ ($b=0$) or $2\coth(\pi h/l)$ ($b=0.5l$). The
	smaller the transverse distance between vortices $h$, the more
	intense is the energy separation for a staggered K\'arm\'an street,
	with the maximum effect $E=\Gamma/(2li_{0\infty})$ being attained
	at $h=0$. For a symmetric wake, on the contrary, as the transverse
	distance decreases, the intensity decreases. The maximum
	efficiency (at $h\rightarrow+\infty$) is equal to the minimum for
	a staggered K\'arm\'an street:
	$E=\Gamma/(2li_{0\infty})\left[1-\Gamma/(4l)\right]$.

	A vivid example of the effect of a staggered wake structure with $N=2$ on $E$ is clearly seen  in Fig.~\ref{fig:I0}a. Upstream and downstream
	from $x\approx 12$, the wake structure is different and $E$ is significantly different. To estimate the parameters, we choose two
	vortices upstream ($6<x<10$) and two vortices downstream ($16<x<18$). Thus, upstream $l\approx3.7$, $h\approx0.29$
	($\Gamma_{0}\approx3.8$), and downstream $l\approx2.8$, $h\approx2.12$ ($\Gamma_{0}\approx2.86$). The calculations of $E$ using
	Eq.~\eqref{eq:E_N=2} give ($i_{0\infty}=129/8$)
	\begin{align*}
	&\text{Upstream: } &&E=0.056~(U=-0.12,~\overline{u'}_{\min}=-1.03),\\
	&\text{Downstream: } &&E=0.032~(U=-0.50,~\overline{u'}_{\min}=-1.02).
	\end{align*}
	The result is that the efficiency decreases, which contradicts the direct numerical simulations in Fig.~\ref{fig:I0}a. This
	can be explained by the following. The point vortex model does not
	describe correctly the velocity fields near the vortex cores. When
	the vortices in the upstream part of the wake are located near
	the centerline, this drawback of the model does not allow one to
	achieve qualitatively correct values of $\overline{u'}_{min}$
	and, hence, correct values of $E$. However, for the downstream part of
	the flow it is not a problem. Indeed,  from the numerical
	results, we have: 
	\begin{align*}
	&\text{Upstream: } &&E=0.013~(U=-0.2,~\overline{u'}_{\min}=-0.3),\\
	&\text{Downstream: } &&E=0.034~(U=-0.4,~\overline{u'}_{\min}=-0.97).
	\end{align*}
	We can see that the main discrepancy with the model is in
	$\overline{u'}_{\min}$ upstream. Downstream
	parameters are in good agreement. It is possible to obtain much
	better agreement in the upstream wake if we replace the point
	vortices in the model by Rankine vortices
	(Eq.~\eqref{eq:PVM_u'_RV}), giving $E\approx0.020$,
	$\overline{u'}_{\min}\approx-0.38$ ($U=-0.12$). An approximate
	value for the vortex core radius was taken equal to 0.5 in accordance with the numerical results.
	
	As shown above, when the Rankine vortices are used to approximate
	the velocity field it is possible to obtain a general expression
	for $\overline{I_0}$ and efficiency $E$. When considering
	Eq.~\eqref{eq:PVM_u'_RV} with $N=2$, one can see that if $h/2$ is
	greater than the vortex core radius $R$, then $E$ is the same as
	for the potential vortices and increases with decrease in $h$.
	However, if $h < 2R$ the value of $E$ decreases to zero with
	$h\rightarrow 0$. Thus, one can expect that in a real flow for
	fixed $\Gamma_0, l, R$ the efficiency of energy separation $E$ has
	its maximum when $h$ is close to $2R$.

	\begin{figure*}[!ht]
		\centering
		\begin{subfigure}[b]{140mm} \centering
			\includegraphics[width=140mm] {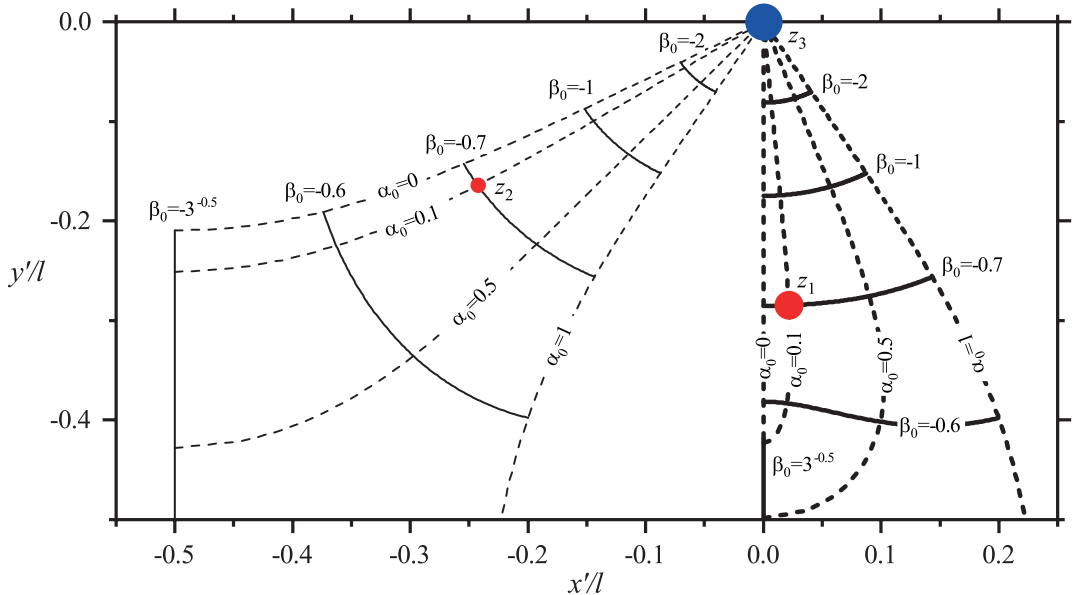}\vspace{5px}
		\end{subfigure}\\
		\caption{The positions of the vortices for different parameters $\alpha_0$ and $\beta_0$. Dashed and solid lines are the isolines of
			$\alpha_0=\text{const}$ and $\beta_0=\text{const}$. Thick and thin lines correspond to the positions of the first and second vortices. Circles
			show an example of a mutual arrangement of vortices for $\alpha_0=0.1$ and $\beta_0=-0.7$.}\label{fig:N=3}
	\end{figure*}

	Let us also compare the numerical results for this case with the
	maximal possible efficiency predicted by the model
	(Eq.~\eqref{eq:Emax}): $E_{\max}\approx 0.032$ upstream and
	downstream (since $\Gamma/l$ is almost equal). This value is close
	to $E$ in the real flow downstream, because the vortex cores have
	a smaller impact on the time-averaged velocity field.
	
	This comparison shows that the point vortex models make it
	possible to estimate values of $E$ if the vortex cores are located
	far enough from the region of minimum values of $\overline{I_0}$.
	In case they are not, one can improve the estimate by the
	replacement of point vortices with Rankine vortices. When vortex
	cores are less involved in the formation of the time-averaged
	velocity field inside the region with minimum values of
	$\overline{I_0}$, the efficiency of energy separation is greater.

	As noted above, for a symmetric vortex street $E$ increases with $h$. That is what we observed in Section~\ref{sec:Calculations}, for example, in Fig.~\ref{fig:I0}d. One can approximate the vortex wake
	fragment in Fig.~\ref{fig:I0}d at $4<x<9$ by a symmetric vortex
	street: from averaging using four vortices $l\approx3.9$,
	$h\approx2.5$ and $\Gamma_0\approx4.2$. From the model
	(Eq.~\eqref{eq:E_N=2}) $E\approx0.03$ which is in qualitative
	agreement with the calculations $E\approx0.026$.
	
	\subsubsection{N=3}
	
	We assume that $\Gamma_1, \Gamma_2\ge0$, $\Gamma_1+\Gamma_2=-\Gamma_3=\Gamma_0=0.5\Gamma>0$, and $\alpha_0=\Gamma_2/\Gamma_1\le1$
	(Fig.~\ref{fig:PVM_Scheme}c). Without loss of generality, we assume  that $z_3=0$ and $-0.5\le x_1'/l< 0.5$, $-0.5\le x_2'/l< 0.5$. We consider
	the equilibrium configurations \cite{StremlerRelativeequilibriasingly2003}, which
	satisfies the relations:
	\begin{align}
	\begin{split}
	\cot\frac{\pi z_1}{l}=\frac{\pm\alpha_0\sqrt{3\beta_0^2-1}-i(\alpha_0+2)\beta_0}{\sqrt{\alpha_0^2+\alpha_0+1}}=a_1+ib_1,\\
	\cot\frac{\pi z_2}{l}=\frac{\mp\sqrt{3\beta_0^2-1}-i(2\alpha_0+1)\beta_0}{\sqrt{\alpha_0^2+\alpha_0+1}}=a_2+ib_2.\end{split}\label{eq:PVM_N3}
	\end{align}
	Here, $\beta_0=lU/S=\beta\Gamma/S$, $S=0.5\Gamma
	\sqrt{\alpha_0^2+\alpha_0+1}/(\alpha_0+1)$. The solution is
	determined by two parameters: $\alpha_0$ and $\beta_0$.
	
	For $3\beta_0^2>1$, different signs in Eqs.~\eqref{eq:PVM_N3}
	correspond to the solutions which are symmetric about $x'=0$. We
	choose `$+$' in the first equation and `$-$' in the second equation.
	Figure ~\ref{fig:N=3} shows the relative positions of the vortices
	for different parameters $\alpha_0$ and $\beta_0$.

	The minimum value of $\overline{u'}_{min}=-\Gamma/(2l)$ is
	attained for $\max(y_1', y_2')<y'<y_3'$, and the efficiency of energy
	separation is
	\begin{equation*}
	E=\frac{\Gamma}{2li_{0\infty}}\left(1+\beta\frac{\Gamma}{l}\right),
	\end{equation*}
	For the considered case
	\begin{equation*}
	\beta=\beta_0\frac{\sqrt{\alpha_0^2+\alpha_0+1}}{2(\alpha_0+1)}<-\frac{1}{4}.
	\end{equation*}
	$\beta\rightarrow-1/4$ as $\beta_0\rightarrow-1/\sqrt{3}$ and $\alpha_0=1$. Thus
	\begin{equation*}
	E<\frac{\Gamma}{2li_{0\infty}}\left(1-\frac{\Gamma}{4l}\right).
	\end{equation*}

	If $3\beta_0^2\le 1$, then the real part of the cotangent is equal
	to $0$, therefore the possible values of $x'_1, x'_2$ are $0$,
	$-l/2$. We consider only the case when the factor $(1+U)$ is
	maximal, i.e. $U=0$ and $\beta=0$. The first and second vortices
	are located on both sides of the axis $y'=0$ (we assume that
	$y'_1<0$). Thus, the efficiency of energy separation is
	\begin{align*}
	E=\frac{\Gamma}{2li_{0\infty}(1+\alpha_0)}.
	\end{align*}
	The value $\alpha_0=0$ corresponds to a staggered vortex street
	with $N=2$ and $h=0$. Otherwise, the scheme  is less efficient.

	\subsubsection{N=4}
	\begin{figure*}[!t]
	\centering
	\begin{subfigure}[b]{80mm} \centering
	\includegraphics[width=80mm] {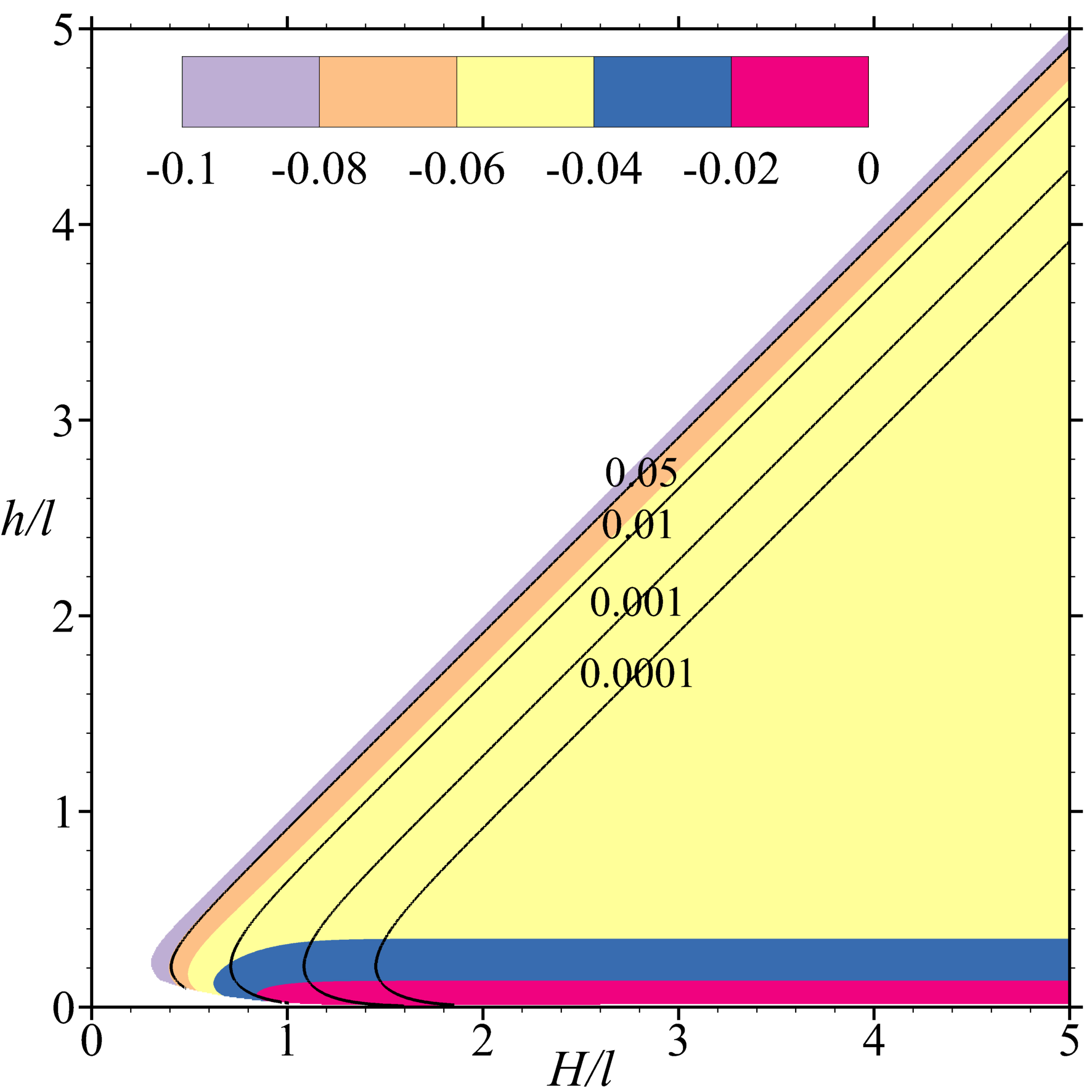}\vspace{5px}
	\caption{$\Gamma_0/l=0.1$}
	\end{subfigure}~
	\begin{subfigure}[b]{80mm} \centering
	\includegraphics[width=80mm] {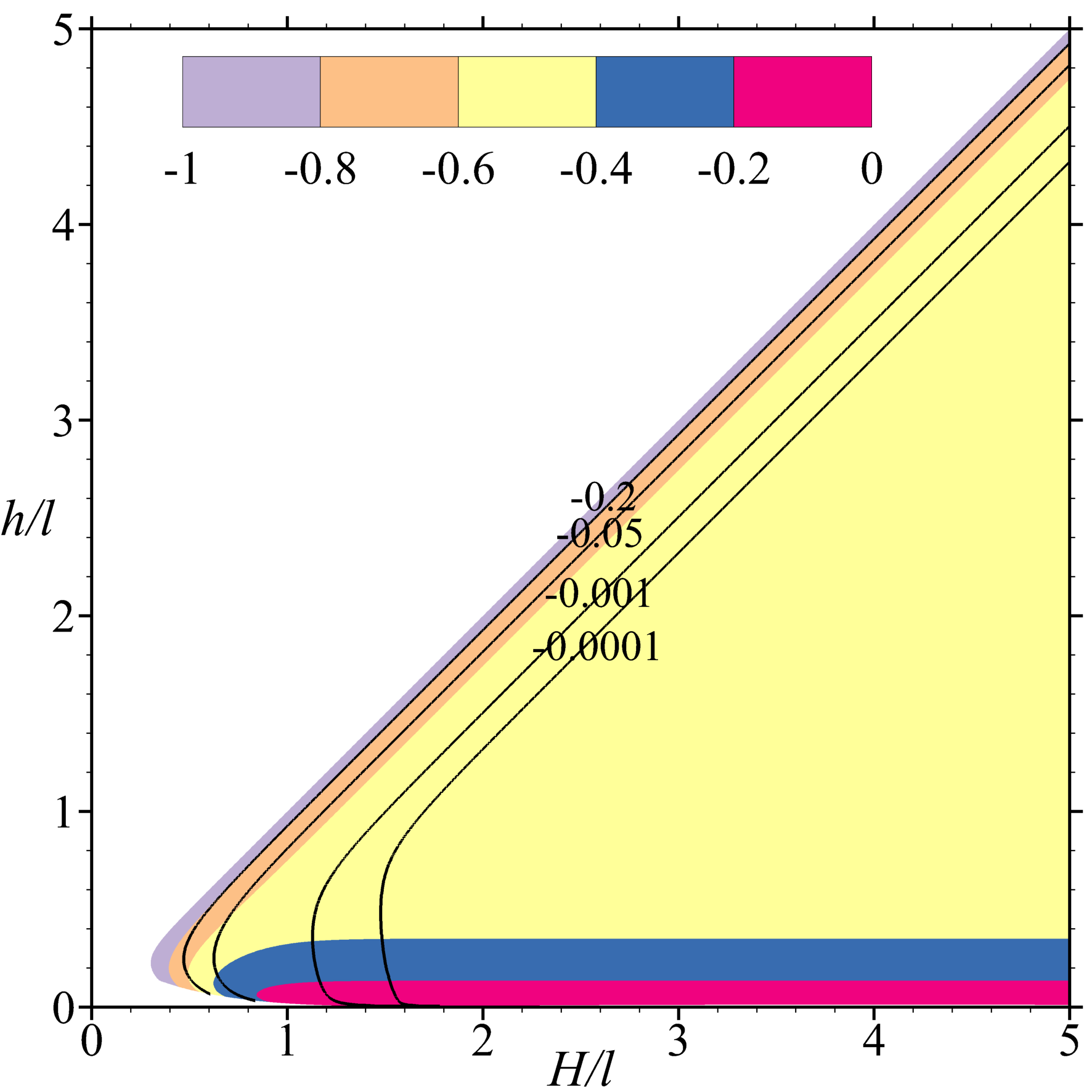}\vspace{5px}
	\caption{$\Gamma_0/l=1$}
	\end{subfigure}\\
	\caption{The values of $U$ (filled plot) at $N=4$ for different
		mutual arrangement of vortices and the difference $\Delta Ei_{0\infty}$ (solid lines) between
		the efficiency of energy separation for $N=4$ and $N=2$. $\Gamma_0/l$ is fixed and equals $0.1$ (left plot) and $1$ (right plot). (For interpretation of the references to color in this figure legend, the reader is referred to the web version of this article.)}\label{fig:N4}
	\end{figure*}
	We consider a symmetric 2P wake  (Fig.~\ref{fig:PVM_Scheme}d),
	which can be observed behind a pair of cylinders. The point vortex models for these
	configurations were studied in \cite{BasuExploringdynamics2P2017}.
	We consider the following case: $\Gamma_1=-\Gamma_4>0$,
	$\Gamma_3=-\Gamma_2>0$, $\Gamma_3<\Gamma_1$  and
	$x'_1+iy'_1=x'_4-iy'_4$, $x'_2+iy'_2=x'_3-iy'_3$,
	$z_2-z_1=-l/2+ih$, $z_3-z_2=i(H-h)$,  $H>h>0$. Thus, $\alpha_1=-\alpha_4>0.25$ and
	$\alpha_3=-\alpha_2<0.25$. Since $2(\alpha_1+\alpha_3)=1$, we can use
	only one parameter $\alpha=2\alpha_3<0.5$, which is equal to
	parameter $\gamma$ in \cite{BasuExploringdynamics2P2017}.

	Let us discuss the possibility of increasing the efficiency of
	energy separation in such configurations, as compared to the
	classical vortex street ($N=2$). To be able to compare the results
	for $N=2$ and $N=4$, we assume that $\Gamma_0$ from
	Section~\ref{sec:N=2} equals $(\Gamma_1+\Gamma_3)/2$. As the
	distance $H/l$ between two vortex streets tends to $+\infty$
	($\alpha\rightarrow0.5$, see \cite{BasuExploringdynamics2P2017})
	$\Gamma_1$ and $\Gamma_3$ tend to $\Gamma_0$, so this assumption
	makes sense.
	
	The minimal value of $\overline{u'}$ is attained for
	$y'_1<y'<y'_2$ and $y'_3<y'<y'_4$. It equals $-\Gamma_1/l$, which
	is not greater than the value $-\Gamma_0/l$ for $N=2$. Although
	$|\overline{u'}_{\min}|$  for $N=4$ is greater than for $N=2$, to
	make a conclusion about efficiency $E$ one should take into
	account the velocities of vortices, i.e. the factor $(1+U)$. In
	Fig.~\ref{fig:N4}, we plotted $-1<U<0$ and lines $\Delta
	E=E_{(N=4)}-E_{(N=2)}=\text{const}$ as functions of $h/l$ and
	$H/l$ at fixed $\Gamma_0/l=0.1$ (Fig.~\ref{fig:N4}a) and $1$
	(Fig.~\ref{fig:N4}b). These examples demonstrate the possibility
	to obtain $\Delta E$ of different signs.  It
	seems particularly interesting that theoretically the creation of
	two vortex streets can improve the efficiency of energy separation
	($\Delta E>0$ at $\Gamma_0/l=0.1$ in Fig.~\ref{fig:N4}a).
	
	\section{Conclusions} In the present study, a clear link between the structure and
	intensity of a vortex street in a wake behind an oscillating
	cylinder and the efficiency of energy separation is demonstrated. Possible ways to improve energy separation efficiency in the wake are investigated. The study is based on both direct numerical simulations of the wake behind a transversely
	oscillating cylinder within the Navier-Stokes equations and the	use of simplified point vortex models of vortex streets with different configurations. 
	
	It is demonstrated that the lowest values of time-averaged total
	enthalpy observed near the centerline of the developed wake can be
	significantly changed by the  variation in the amplitude and
	frequency of forced oscillations. Pronounced changes in the area of
	the region of reduced total enthalpy in the wake can also be
	achieved. These effects are attributable to the formation of
	qualitatively different vortex structures in the wake. The
	appearance of reversed vortex pairs (with the positive vortex
	located above the negative one) leads to the development of
	regions with increased total enthalpy.

	The velocity fields for different vortex street structures  were
	approximated by point vortex models, consisting of 2, 3, and 4
	infinite periodic vortex chains in the equilibrium state. Based on
	this approximation of the velocity fields, the distribution of total
	enthalpy was obtained. This approach makes it possible to
	easily estimate the maximum efficiency of energy separation due to
	pressure oscillations in the wake. For instance, in the general case
	of $N$ vortex chains in the street for any fixed mutual
	arrangement of vortices the maximum possible efficiency is equal
	to $\Gamma(N-1)/(2l i_{0\infty}N)$ for relative velocity of vortices
	equal to $0.5$. The fixed arrangement here means the freezing of
	vortex locations (with respect to a characteristic length $l$), which takes place at constant normalized vortex circulations $\alpha_k$ and coefficient $\beta=Ul/\Gamma$.
	
	It is demonstrated that the point vortex models can predict
	the energy separation efficiency well if the motion of vortex cores
	occurs outside the region considered. Otherwise, the results of
	the model can be improved using the Rankine vortices  instead of
	the potential ones, since the main discrepancy is caused by an
	incorrect velocity distributions inside the vortex cores. Using
	the improved model for a staggered K\'arm\'an street, it was shown
	that the maximum efficiency of energy separation can be achieved
	when the transverse distance between opposite vortices is close to
	the vortex core diameter. This model was successfully used to
	simulate the effect of a significant increase in  energy separation
	efficiency due to the natural restructuring of the vortex street
	in the wake behind a fixed cylinder.
	
	For a symmetric K\'arm\'an street, the efficiency is restricted by
	the minimum value of efficiency for the staggered arrangement. In
	contrast to the staggered arrangement, the efficiency increases
	with the increase in the transverse distance between vortices.
	
	The theoretical results show the possibility  to increase the
	efficiency of energy separation by creating side-by-side
	K\'arm\'an vortex streets. This configuration can be formed, for
	example, behind a pair of cylinders placed far enough from each
	other.
	
	It should be noted that all the results are obtained using two-dimensional models. For a quantitative description of energy separation in the real flows one should take into account three-dimensional effects, turbulence, and, probably, the dependence of the viscosity and thermal conduction coefficients on temperature. Nevertheless, the present study provides preliminary estimate of the influence of particular vortex structures on energy separation due to the pressure variation mechanism. 
	
	\section*{Acknowledgements}
	The work was supported by the Russian Science Foundation (project 14-19-00699). The research was carried out using the equipment of the shared research facilities of HPC computing resources at Lomonosov Moscow State University.
	
	\appendix
	\section{The Navier--Stokes equations in primitive variables}\label{appA}
	The matrices in Eq.~\eqref{eq:NS} take the following form:
	
	\begin{align*}
	\mat{A}_0=\frac{1}{(\gamma-1)T}\left(
	\begin{array}{cccc}
	1 & 0 & 0 & -\frac{p}{T} \\
	u_1 & p & 0 & -\frac{pu_1}{T} \\
	u_2 & 0 & p & -\frac{pu_2}{T} \\
	\varepsilon & pu_1 & pu_2 & -\frac{p|\vec{u}|^2}{2T}
	\end{array}
	\right),
	\end{align*}
	\begin{align*}
	\mat{A}_i=\frac{1}{(\gamma-1)T}\left(
	\begin{array}{cccc}
	u_i & p\delta_{1i} & p\delta_{2i} & -\frac{pu_i}{T} \\
	u_iu_1 & pu_i(1+\delta_{1i}) & pu_1\delta_{2i} & -\frac{pu_iu_1}{T} \\
	u_iu_2 & pu_2\delta_{1i} & pu_i(1+\delta_{2i}) & -\frac{pu_iu_2}{T} \\
	u_i\varepsilon & pu_iu_1+\delta_{1i}p\varepsilon & pu_iu_2+\delta_{2i}p\varepsilon & -\frac{pu_i|\vec{u}|^2}{2T}
	\end{array}
	\right),
	\end{align*}
	\begin{align*}
	\scriptstyle
	\mat{K}_{11}=\left(
	\begin{array}{cccc}
	0 & 0 & 0 & 0 \\
	0 & \frac{4}{3\Rey} & 0 & 0 \\
	0 & 0 & \frac{1}{\Rey} & 0 \\
	0 & \frac{4u_1}{3\Rey} & \frac{u_2}{\Rey} & \frac{\gamma}{\Pran\Rey} \\
	\end{array}
	\right),\,
	\mat{K}_{12}=\left(
	\begin{array}{cccc}
	0 & 0 & 0 & 0 \\
	0 & 0 & -\frac{2}{3\Rey} & 0 \\
	0 & \frac{1}{\Rey} & 0 & 0 \\
	0 & \frac{u_2}{\Rey} & -\frac{2u_1}{3\Rey} & 0
	\end{array}
	\right),
	\end{align*}
	\begin{align*}
	\scriptstyle
	\mat{K}_{21}=\left(
	\begin{array}{cccc}
	0 & 0 & 0 & 0 \\
	0 & 0 & \frac{1}{\Rey} & 0 \\
	0 & -\frac{2}{3\Rey} & 0 & 0 \\
	0 & -\frac{2u_2}{3\Rey} & \frac{u_1}{\Rey} & 0 \\
	\end{array}
	\right),\,
	\mat{K}_{22}=\left(
	\begin{array}{cccc}
	0 & 0 & 0 & 0 \\
	0 & \frac{1}{\Rey} & 0 & 0 \\
	0 & 0 & \frac{4}{3\Rey} & 0 \\
	0 & \frac{u_1}{\Rey} & \frac{4u_2}{3\Rey} & \frac{\gamma}{\Pran\Rey}
	\end{array}
	\right),
	\end{align*}
	\begin{align*}
	\vec{P}_i=p\left(0, \delta_{1i}, \delta_{2i}, u_i\right)^*,\
	\vec{R}=-\rho v_{c,t}\left(0, 0, 1, u_2\right)^*.
	\end{align*}
	Here, $\varepsilon=T+0.5|\vec{u}|^2$, $\delta_{ij}$ is Kronecker delta, $i=1,2$, $j=1,2$.
	
	\bibliographystyle{elsarticle-num}
	\bibliography{Aleksyuk}

\begin{thebibliography}{10}
\expandafter\ifx\csname url\endcsname\relax
  \def\url#1{\texttt{#1}}\fi
\expandafter\ifx\csname urlprefix\endcsname\relax\def\urlprefix{URL }\fi
\expandafter\ifx\csname href\endcsname\relax
  \def\href#1#2{#2} \def\path#1{#1}\fi

\bibitem{KurosakaEnergyseparationvortex1987}
M.~Kurosaka, J.~B. Gertz, J.~E. Graham, J.~R. Goodman, P.~Sundaram, W.~C.
  Riner, H.~Kuroda, W.~L. Hankey, Energy separation in a vortex street, Journal
  of Fluid Mechanics 178 (1987) 1--29.
\newblock \href {http://dx.doi.org/10.1017/S0022112087001095}
  {\path{doi:10.1017/S0022112087001095}}.

\bibitem{RyanExperimentsaerodynamiccooling1951}
L.~F. Ryan, Experiments on aerodynamic cooling, Ph.D. thesis, Swiss Federal
  Institute of Technology, Zurich (1951).

\bibitem{ThomannMeasurementsRecoveryTemperature1959}
H.~Thomann, Measurements of the {{Recovery Temperature}} in the {{Wake}} of a
  {{Cylinder}} and of a {{Wedge}} at {{Mach Numbers Between}} 0. 5 and 3, Tech.
  Rep. Report 84, {National Aeronautical Research Institute (FFA)}, Sweden
  (1959).

\bibitem{EckertCrosstransportenergy1987}
E.~R.~G. Eckert, Cross transport of energy in fluid streams, W\"arme-und
  Stoff\"ubertragung 21~(2-3) (1987) 73--81.
\newblock \href {http://dx.doi.org/10.1007/BF01377562}
  {\path{doi:10.1007/BF01377562}}.

\bibitem{NgTimeresolvedmeasurementstotal1990}
W.~F. Ng, W.~M. Chakroun, M.~Kurosaka, Time-resolved measurements of total
  temperature and pressure in the vortex street behind a cylinder, Physics of
  Fluids A: Fluid Dynamics 2~(6) (1990) 971--978.
\newblock \href {http://dx.doi.org/10.1063/1.857604}
  {\path{doi:10.1063/1.857604}}.

\bibitem{KulkarniEnergyseparationwake2009}
K.~Kulkarni, R.~Goldstein, Energy separation in the wake of a cylinder:
  {{Effect}} of {{Reynolds}} number and acoustic resonance, International
  Journal of Heat and Mass Transfer 52~(17-18) (2009) 3994--4000.
\newblock \href {http://dx.doi.org/10.1016/j.ijheatmasstransfer.2009.03.024}
  {\path{doi:10.1016/j.ijheatmasstransfer.2009.03.024}}.

\bibitem{AleksyukDirectnumericalsimulation2018}
A.~I. Aleksyuk, A.~N. Osiptsov, Direct numerical simulation of energy
  separation effect in the near wake behind a circular cylinder, Int. J. Heat
  Mass Transfer 119 (2018) 665--677.
\newblock \href {http://dx.doi.org/10.1016/j.ijheatmasstransfer.2017.11.133}
  {\path{doi:10.1016/j.ijheatmasstransfer.2017.11.133}}.

\bibitem{Eiamsa-ardReviewRanqueHilsch2008}
S.~{Eiamsa-ard}, P.~Promvonge, Review of {{Ranque}}\textendash{{Hilsch}}
  effects in vortex tubes, Renewable and sustainable energy reviews 12~(7)
  (2008) 1822--1842.
\newblock \href {http://dx.doi.org/10.1016/j.rser.2007.03.006.}
  {\path{doi:10.1016/j.rser.2007.03.006.}}

\bibitem{LeontevGasdynamicmethod1997}
A.~I. Leont'ev, Gas\textendash{}dynamic method of energy separation of gas
  flows, High Temp. 35~(1) (1997) 155--157.

\bibitem{LeontievExperimentalinvestigationmachinefree2017}
A.~I. Leontiev, A.~G. Zditovets, Y.~A. Vinogradov, M.~M. Strongin, N.~A.
  Kiselev, Experimental investigation of the machine-free method of temperature
  separation of air flows based on the energy separation effect in a
  compressible boundary layer, Experimental Thermal and Fluid Science 88 (2017)
  202--219.
\newblock \href {http://dx.doi.org/10.1016/j.expthermflusci.2017.05.021}
  {\path{doi:10.1016/j.expthermflusci.2017.05.021}}.

\bibitem{EckertMessungenTemperaturverteilungauf1942}
E.~Eckert, W.~Weise, Messungen der {{Temperaturverteilung}} auf der
  {{Oberfl\"ache}} schnell angestr\"omter unbeheizter {{K\"orper}}, Forschung
  im Ingenieurwesen 13~(6) (1942) 246--254.
\newblock \href {http://dx.doi.org/10.1007/BF02585343}
  {\path{doi:10.1007/BF02585343}}.

\bibitem{goldstein_energy_2008}
R.~J. Goldstein, K.~S. Kulkarni, Energy {{Separation}} in the {{Wake}} of a
  {{Cylinder}}, J. Heat Transfer 130~(6) (2008) 061703--061703--9.
\newblock \href {http://dx.doi.org/10.1115/1.2891222}
  {\path{doi:10.1115/1.2891222}}.

\bibitem{WilliamsonVortexformationwake1988}
C.~Williamson, A.~Roshko, Vortex formation in the wake of an oscillating
  cylinder, Journal of Fluids and Structures 2~(4) (1988) 355--381.
\newblock \href {http://dx.doi.org/10.1016/S0889-9746(88)90058-8}
  {\path{doi:10.1016/S0889-9746(88)90058-8}}.

\bibitem{LeontiniWakestateenergy2006}
J.~S. Leontini, B.~E. Stewart, M.~C. Thompson, K.~Hourigan, Wake state and
  energy transitions of an oscillating cylinder at low {{Reynolds}} number,
  Physics of Fluids 18~(6) (2006) 067101.
\newblock \href {http://dx.doi.org/10.1063/1.2204632}
  {\path{doi:10.1063/1.2204632}}.

\bibitem{KochinTheoreticalhydromechanics1964}
N.~E. Kochin, I.~A. Kibel, N.~V. Roze, Theoretical Hydromechanics,
  {Interscience}, 1964.

\bibitem{MeleshkoDynamicsvortexstructures1993}
V.~V. Meleshko, M.~Y. Konstantinov, Dynamics of Vortex Structures, {Naukova
  Dumka, Kiev}, 1993.

\bibitem{Arefmotionthreepoint1996}
H.~Aref, M.~A. Stremler, On the motion of three point vortices in a periodic
  strip, Journal of Fluid Mechanics 314 (1996) 1--25.
\newblock \href {http://dx.doi.org/10.1017/S0022112096000213}
  {\path{doi:10.1017/S0022112096000213}}.

\bibitem{BasuExploringdynamics2P2017}
S.~Basu, M.~A. Stremler, Exploring the dynamics of `{{2P}}' wakes with
  reflective symmetry using point vortices, Journal of Fluid Mechanics 831
  (2017) 72--100.
\newblock \href {http://dx.doi.org/10.1017/jfm.2017.563}
  {\path{doi:10.1017/jfm.2017.563}}.

\bibitem{StremlerRelativeequilibriasingly2003}
M.~A. Stremler, Relative equilibria of singly periodic point vortex arrays,
  Physics of Fluids 15~(12) (2003) 3767--3775.
\newblock \href {http://dx.doi.org/10.1063/1.1624608}
  {\path{doi:10.1063/1.1624608}}.

\bibitem{AleksyukAnalysisthreedimensionaltransition2018}
A.~I. Aleksyuk, V.~Y. Shkadov, Analysis of three-dimensional transition
  mechanisms in the near wake behind a circular cylinder, European Journal of
  Mechanics - B/Fluids 72 (2018) 456--466.
\newblock \href {http://dx.doi.org/10.1016/j.euromechflu.2018.07.011}
  {\path{doi:10.1016/j.euromechflu.2018.07.011}}.

\bibitem{AleksyukFormationevolutiondecay2012}
A.~I. Aleksyuk, V.~P. Shkadova, V.~Y. Shkadov, Formation, evolution, and decay
  of a vortex street in the wake of a streamlined body, Moscow University
  Mechanics Bulletin 67~(3) (2012) 53--61.
\newblock \href {http://dx.doi.org/10.3103/S0027133012030016}
  {\path{doi:10.3103/S0027133012030016}}.

\bibitem{Blackburnstudytwodimensionalflow1999}
H.~M. Blackburn, R.~D. Henderson, A study of two-dimensional flow past an
  oscillating cylinder, Journal of Fluid Mechanics 385 (1999) 255--286.
\newblock \href {http://dx.doi.org/10.1017/S0022112099004309}
  {\path{doi:10.1017/S0022112099004309}}.

\bibitem{CrowdySpeedKarmanpoint2017}
D.~G. Crowdy, V.~S. Krishnamurthy, Speed of a von {{K\'arm\'an}} point vortex
  street in a weakly compressible fluid, Physical Review Fluids 2~(11) (2017)
  114701.
\newblock \href {http://dx.doi.org/10.1103/PhysRevFluids.2.114701}
  {\path{doi:10.1103/PhysRevFluids.2.114701}}.

\end{thebibliography}
	
	\end{document}